\documentclass[
    10pt
    ,aps
    ,pop%,rmp%,pra%,pop
    ,oneside
    ,nobibnotes
    ,floatfix
    ,showpacs
    ,twocolumn
    ,nofootinbib
]{revtex4-2}

\usepackage{amsmath,amssymb,bm}
\allowdisplaybreaks[1]
\usepackage{siunitx}
\NewDocumentCommand{\bnum}{O{}m}{%
  \num[math-rm=\mathbf,#1]{#2}%
}

\usepackage{graphicx}\graphicspath{{./}{../}}

\usepackage{tabularx}

\usepackage{iftex}
\iftutex
    \usepackage{fontspec}
    \usepackage[math-style=ISO,bold-style=ISO]{unicode-math}
\else
    \usepackage[T1]{fontenc} % UTF-8: PDFLaTeX
    \usepackage[utf8]{inputenc} % UTF-8: PDFLaTeX
    \usepackage{txfonts} % UTF-8: PDFLaTeX
\fi

\usepackage[english, showlanguages]{babel}

    \iftutex
        \renewcommand{\vec}[1]{\symbf{#1}}
    \else
        \renewcommand{\vec}[1]{\mathbf{#1}}
    \fi

    \DeclareMathOperator{\e}{e}
    \DeclareMathOperator{\const}{const}

    \newcommand{\parder}[2]{\frac{\partial #1}{\partial #2}}

    \newcommand{\dif}[2][]{\mathop{}\!\mathrm{d}
        \if
            \relax\detokenize{#1}\relax
        \else
            ^{\mkern-1.mu#1}\mkern-2.5mu %[S2]
    \fi
    #2\,}
    \newcommand{\der}[2]{\frac{\dif{#1}}{\dif{#2}}}
    
    \usepackage[svgnames]{xcolor}
    \definecolor{darkblue}{cmyk}{1.00, 0.50, 0.00, 0.40}

\usepackage[
    colorlinks
        ,linkcolor=violet
        ,filecolor=purple
        ,citecolor=teal
        ,urlcolor=magenta
    ,unicode
    ,pdfpagemode=UseOutlines% (show bookmarks)
    ,pdfdisplaydoctitle=true% display document title instead of file
    ,pdfusetitle
]{hyperref}
    \hypersetup{
        ,pdfkeywords={GDMT, ballooning modes, large-m limit}
    }
\usepackage{bookmark}

\begin{document}

\def\bot{\mathrel\perp}

\selectlanguage{english}

\title{On the stability of small-scale ballooning modes in mirror traps}
\author{Igor Kotelnikov}
    \email{I.A.Kotelnikov@inp.nsk.su}
    \affiliation{Budker Institute of Nuclear Physics SB RAS, Novosibirsk, 630090, Russia}
    \affiliation{Novosibirsk State University, Novosibirsk, 630090, Russia}
%\author{Dmitry Yakovlev}
%    %\email{D.V.Yakovlev@inp.nsk.su}
%    \affiliation{Budker Institute of Nuclear Physics SB RAS, Novosibirsk, 630090, Russia}
%    \affiliation{Novosibirsk State University, Novosibirsk, 630090, Russia}
\author{Qiusun Zeng}
    %\email{Qiusun.Zeng@fds.org.cn}
    %\email{Qiusun.Zeng@inest.cas.cn}
    \affiliation{Institute of Nuclear Energy Safety Technology HFIPS CAS, Hefei, China}

\date{\today}
%\pacs{
%    % ==========================
%    %http://publish.aps.org/PACS
%    % ==========================
%    %52.50.Qt; %	Plasma heating by radio-frequency fields; ICR, ICP, helicons
%    %52.50.Sw; %	Plasma heating by microwaves; ECR, LH, collisional heating
%}

\begin{abstract}
    %This Note \#02 was started on Wednesday, April 8, 2020.
    %This is internal document not intended for publications. Any comments, criticisms are welcome. Part of the material was borrowed from the MHD textbook, which is being prepared for publication, so the style of presentation is too “flamboyant” in some places.
    %
    %The calculations described here were performed by Wolfram Mathematica in the \texttt{214 - Ballooning Instability - beta crit.nb} notebook.
    %
    It is shown that steepening of the radial plasma pressure profile leads to a decrease in the critical value of beta, above which small-scale balloon-type perturbations in a mirror trap become unstable. This means that small-scale ballooning instability leads to a smoothing of the radial plasma profile. This fact seems to have received little attention in the available publications.
    The critical beta values for the real magnetic field of the gas-dynamic trap was also calculated. In the best configuration the critical beta \num[round-precision=2]{0.718989} is obtained for a plasma with a parabolic radial pressure profile.
\end{abstract}

\maketitle

\section{Introduction}\label{s1}

%В научной литературе имеется некоторое число публикаций, в той или иной мере имеющих отношение к теории устойчивости МГД-возмущения баллонного типа в открытых ловушках
There are a number of publications, in one way or another, related to the stability of ballooning MHD perturbations in open traps
\cite{%
    %==========================================
    % High-m
    RyutovStupakov1980BINP_130e,%m=\infty
    RyutovStupakov1981IAEA_1_119,%m=\infty
    BushkovaMirnov1985BINP_103e,%m=\infty
    BushkovaMirnov1986VANT_2_19e,%m=\infty
    Newcomb1981JPP_26_529,% paraxial, общая теория без КЛР
    %DIppolitoHafizi1981PF_24_2274,% Low-m, axisymmetric, sharp-boundary; SEE BELOW
    DIppolitoMyra1981PF_24_2265,% large-m, axisymmetric, ур-е как у Бушковой-Мирнова
    DIppolitoMyraOgden1982NF_24_707,% High-m ballooning stability of an axisymmetric e-ring-stabilized tandem mirror
    DIppolitoHafiziMyra1982PF_25_2223,% large-m; Low-m sharp-boundary; axisymmetric; вывод ур-я как у Бушковой-Мирнова
    Bernstein+1958ProcRSos_17_244,% энергетический принцип и непараксиальный случай
    %=========================================
    %Horton1981PF_24_1270,% Drift modes in axisymmetric tandem mirrors
    KaiserPearlstein1983PhysFluids_26_3053,% Quadrupole mirror; derivation of ballobing eqs
    CloseLichtenberg1989PFB_1_629, %  Experiment on MMX at Berkly
    Tsidulko1992BINP_92-10e,% Резистивная баллонная мода
    BilikmenMirnovOke1997NF_37_973,% Localized Ballooning Modes In Two Component Gas Dynamic Trap
    %=========================================
    % Non-paraxial
    Arsenin1983JETPhLett_37_637,% \beta \ll 1; Non-Paraxial Equilibria;
    %%Arsenin1999FT_35_3,% минибзор; Non-Paraxial Equilibria; wall stabilization
    %%ArseninKuyanov2001PPR_27_635,% Kruskal–Oberman criterion; Divertor; \beta\ll1
    ArseninKuyanov2001FT_39_175,% Non-Paraxial Equilibria; Grad-Shafranov; как у Берштейна
    ArseninZvonkovSkovoroda2005PPR_31_3,% Stabilization of Ballooning Modes by Nonparaxial Cells; periodic system
    Arsenin2008PPR_34_349,% Kruskal–Oberman criterion; periodic trap, m=\infty
    ArseninTerekhin2008PPR_34_895,% Kruskal–Oberman; поле как у Берштейна, m=\infty
    ArseninTerekhin2011PPR_37_723,% Plasma Equilibrium in Flux Coordinates как у Берштейна
    %=========================================
    Rosenbluth+1962NFSuppl_1_143,% Эффекты КЛР
    RobertsTaylor1962PhysRevLett_8_197,% Эффекты КЛР
    Rudakov1962NF_2_107,% Эффекты КЛР
    %TangCatto1981PF_24_1314,% вывод уравнения с КЛР, no wall
    %дLeeCatto1981PF_24_2010, % вывод уравнения с КЛР, no wall
    DIppolitoFrancisMyraTang1981PF_24_2270,% FLR, axisymmetric, any m, критерий m>a^{2}/(\rho L), no wall
    %=========================================
    KaiserNevinsPearlstein1983PF_26_351, %% quadrupole, FLR, m=1
    %=========================================
    % Wall Stabilization
    Berk+1985PPCNFR_2_321,% drift-modes, AIC, быстрые эл., Wall stabilization
    Berk+1984PF_27_2705, % wall stabilizatio, sharp boundary
    %HaasWesson1967PF_10_2245, % sharp boundary, theta-pinch, m=1; не заметили эффект стеночной стабилизации (?)
    DIppolitoHafizi1981PF_24_2274, % sharp boundary, lom m, wall; не заметили эффект стеночной стабилизации (?)
    DIppolitoMyra1984PF_27_2256,% any m, sharp boundary hollow profile, wall stabilizxation, 2nd stability zone; RF stabilization
    %=========================================
    % Учет КЛР & Wall Stabilization
    KaiserPearlstein1985PhysFluids_28_1003,% m=1, eq. for arb. radial profile, FLR included; sharp boundary case;
    Kesner1985NF_25_275,% m=1, sharp boundary, parabolic axial profile of B; wall stabilization
    LiKesnerLane1985NF_25_907, % m=1, wall stabilization; CONDUCTING-WALL AND PRESSURE PROFILE EFFECT ON MHD STABILIZATION OF AXISYMMETRIC MIRROR
    LiKesnerLane1987NF_27_101,% partially enclosed wall. m=1; stepwise
    LiKesnerLoDestro1987NF_27_1259,% m=1, wall stabilization, sharp boundary;
    LoDestro1986PF_29_2329,% Вывод уравнения для жесткой моды при любом рад. профиле
    ArseninKuyanov1996PPR_22_638%,% m=1 axisymmetric; ссылка на Лодестро
}.
Most of them were published in the 80s of the last century. In the next decades, interest in the problem of ballooning instability in open traps significantly weakened (in contrast to what is happening in tokamaks, see e.g.\ \cite{Snyder+2002PoP_9_2037, Halpern+2013PoP_20_052306, Eich+2018NF_58_034001}), which was a consequence of the termination of the TMX (Tandem Mirror eXperiment) and MFTF-B (Mirror Fusion Test Facility B) projects in the USA in 1986 \cite{Ongena+2016NaturePhysics_12_398}. However, the achievement of a high electron temperature and high beta ($\beta$, the ratio of the plasma pressure to the magnetic field pressure) in the Gas-Dynamic Trap (GDT) at the Budker Institute of Nuclear Physics in Novosibirsk \cite{
    Ivanov+PhysRevLett_90_105002, % \beta=0.4
    Simonen+2010JFE_29_558, % \beta=0.6
    Bagryansky+2011FST_59_31,
    Bagryansky+2015PhysRevLett_114_205001,
    Bagryansky+2015NF_55_053009,
    Bagryansky+2016AIPConfProc_030015,
    Bagryansky+2016AIPCP_1771_020003,
    Yakovlev+2018NF_58_094001,
    Bagryansky+2019JFE_38_162}
and the emergence of new ideas \cite{Beklemishev2016PoP_23_082506} and new projects \cite{Granetzny+2018APS_CP11_150, Bagryansky+2020NuclFusion_60_036005} makes us rethink old results.

%{\Russian
%    Повторяя логику прежних исследований, мы сначала обратимся к изучению устойчивости мелкомасштабных баллонных возмущений, которые характеризуются большими значениями азимутального волнового числа $m\gg1$. В первой статье мы ограничимся параксиальным приближением, предполагая, что поперечный размер плазмы $a$ в ловушке мал по сравнению с её длиной $L$, $a\ll L$. Исследованию баллонной неустойчивости в непараксиальной открытой ловушке будет посвящена следующая статья. Ниже мы будем использовать канонический метод исследования баллонной неустойчивости, исторически связанный с энергетическим принципом в приближении одножидкостной магнитной гидродинамики \cite{Bernstein+1958ProcRSos_17_244}.
%}
We are planning to prepare a series of three articles and, following the logic of previous research, in the first article we turn to the study of the stability of small-scale ballooning disturbances, which are characterized by large values of the azimuthal wave number $m$, $m \gg 1$. In this first article, we restrict ourselves to the paraxial approximation, assuming that the transverse size $a$ of the plasma in the trap is small compared to its length $L$, $a \ll L$.
%The next article will be devoted to the study of small-scale ballooning instability in a non-paraxial open trap, and the third article will examine the stability of the rigid ballooning mode $m = 1$.
%
In the next article, the effects of non-paraxiality will be taken into account, and in the third article, the stability of the rigid ballooning mode $m = 1$ will be examined.

We will use the canonical method for studying ballooning instability, historically associated with the energy principle in the approximation of one-fluid ideal magnetohydrodynamics \cite{Bernstein+1958ProcRSos_17_244}. It should be understood that this method has limited applicability. In particular, it assumes that in equilibrium the plasma is stationary, that is, it does not rotate around its axis and does not flows out through magnetic mirrors. In addition, the energy principle ignores the effects of a finite Larmor radius (FLR effects) \cite{
        Rosenbluth+1962NFSuppl_1_143,% Эффекты КЛР
        RobertsTaylor1962PhysRevLett_8_197,% Эффекты КЛР
        Rudakov1962NF_2_107%,% Эффекты КЛР
}, which should stabilize small-scale flute and ballooning perturbations with $m>a^{2}/(L\rho_{i})$ ($\rho_{i}$ is the ion Larmor radius) \cite{DIppolitoFrancisMyraTang1981PF_24_2270}. Nevertheless, even with the listed limitations, the energy principle allows important conclusions to be drawn. In particular, this article will show that small-scale ballooning instability leads to a smoothing of the radial plasma pressure profile.

First part of the study undertaken in the present paper is based on a simplified derivation of the equation for ballooning oscillations from the energy principle performed by Dmitri Ryutov and Gennady Stupakov \cite{RyutovStupakov1980BINP_130e, RyutovStupakov1981IAEA_1_119}. The simplification includes (a) the use of the paraxial approximation, and (b) the use of the low beta approximation, $\beta\ll1$, which allows us to assume that the magnetic field both inside and outside the plasma column is close to the vacuum one.
%With some improvements, this derivation is included in the second part of the textbook ``Lectures on Plasma Physics'' \cite{Kotelnikov2020V2e}, which is currently being prepared for publication.

%Более точный расчёт, который использовал только первое (параксиальное) приближение (а), был сделан в работе О.А.",Бушковой и В.В.",Мирнова \cite{BushkovaMirnov1985BINP_103, BushkovaMirnov1986VANT_2_19}. Эти авторы использовали уравнение баллонных колебаний, которое было получено в У.",Ньюкомбом \cite{Newcomb1981JPP_26_529}. Такое же уравнение с кратким выводом приведено в работе \cite{DIppolitoHafiziMyra1982PF_25_2223} для случая изотропной плазмы.  В работе \cite{KaiserPearlstein1985PhysFluids_28_1003} со ссылками на более ранние работы приведено уравнение для случая анизотропной плазмы.

A more accurate calculation, which used only the first (paraxial) approximation (a), was made by Ol'ga Bushkova and Vladimir Mirnov \cite{BushkovaMirnov1985BINP_103e, BushkovaMirnov1986VANT_2_19e}. These authors used the balloon equation that was derived by William Newcomb in \cite{Newcomb1981JPP_26_529}. The same equation with a short derivation is given in \cite{DIppolitoHafiziMyra1982PF_25_2223} for the case of isotropic plasma.
%%
%% Нет там мелкого маштаба.
%%
%In \cite{KaiserPearlstein1985PhysFluids_28_1003} with references for earlier works, a similar equation is given for the case of anisotropic plasma.
%
It has been studied in \cite{DIppolitoMyra1981PF_24_2265}, and the effect of strong anisotropy created by a rigid ring of electrons has been included into consideration in \cite{DIppolitoMyraOgden1982NF_24_707}.

%В классической работе \cite{Bernstein+1958ProcRSos_17_244} И.",Бернштейна с соавторами, где в современном виде был сформулирован энергетический принцип, в последнем разделе приведено сравнительно простое выражение для потенциальной энергии желобковых и баллонных возмущений в непараксиальной аксиально-симметричной открытой ловушке в приближении бесконечного азимутального числа, когда возмущение представляет собой подобие лезвия ножа. Примеры расчёта непараксиального равновесия плазмы в переменных, которые использованы Бернштейном с соавторами, описаны в моей заметке \cite{DBubble-Note03}. Результаты вычисления предельного бета для мелкомасштабных баллонных колебаний в непараксиальной ловушки описаны в следующей по номеру заметке \cite{DBubble-Note04}. Попытки расчётов равновесия и устойчивости непараксиальных систем  предпринимал В.В.",Арсенин с соавторами
%
In the classical paper \cite{Bernstein+1958ProcRSos_17_244} by Ira Bernstein et al, where the energy principle was formulated in its modern form, in the last section a relatively simple expression for the potential energy of flute and ballooning perturbations in a non-paraxial axially symmetric open trap is given in the approximation of infinite azimuthal number $m\to\infty $, when the disturbance is like a knife blade.
%
%Examples of calculating the non-paraxial equilibrium of plasma in variables, which were used by Bernstein et al., are described in my next note \cite{DBubble-Note03e}. The results of calculating the critical beta for small-scale ballooning oscillations in a non-paraxial trap are described in the next note \cite{DBubble-Note04e}.
%
Attempts to calculate the equilibrium and stability of nonparaxial systems were made by Vladimir Arsenin et al.
\cite{%
    Arsenin1983JETPhLett_37_637,% \beta \ll 1; Non-Paraxial Equilibria;
    %%Arsenin1999FT_35_3,% минибзор; Non-Paraxial Equilibria; wall stabilization
    %%ArseninKuyanov2001PPR_27_635,% Kruskal–Oberman criterion; Divertor; \beta\ll1
    ArseninKuyanov2001FT_39_175,% Non-Paraxial Equilibria; Grad-Shafranov; как у Берштейна
    ArseninZvonkovSkovoroda2005PPR_31_3,% Stabilization of Ballooning Modes by Nonparaxial Cells; periodic system
    Arsenin2008PPR_34_349,% Kruskal–Oberman criterion; periodic trap
    ArseninTerekhin2008PPR_34_895,% Kruskal–Oberman; поле как у Берштейна
    ArseninTerekhin2011PPR_37_723%,% Plasma Equilibrium in in Flux Coordinates как у Берштейна
    %%ArseninKuyanov1996PPR_22_638,% m=1 axisymmetric; ссылка на Лодестро
},
%однако существенного прогресса, как кажется, не достиг.
however, no significant progress seems to have been made.

It is commonly believed that FLR effects
\cite{%
    Rosenbluth+1962NFSuppl_1_143,% Эффекты КЛР
    RobertsTaylor1962PhysRevLett_8_197,% Эффекты КЛР
    Rudakov1962NF_2_107%,% Эффекты КЛР
}
are capable to stabilize all modes with the azimuthal numbers $ m \geq2 $ and impose on the $ m = 1 $ mode the shape of a rigid (``solid-state'') displacement without deformation of the plasma density profile in each section of the plasma column. More accurate treatment is given in \cite{DIppolitoFrancisMyraTang1981PF_24_2270}. It leads the conclusion that ballooning modes with $m> a^{2}/L\rho_{i}$
%(where $\rho_{i}$ is average Larmor radius of ions)
are stable for any $\beta $.
%
%In the same note, publications are cited where such rigid perturbations are investigated or the corresponding equations are written.
%
The stability of ballooning perturbations with $m = 1$ was considered in Refs.~\cite{%
    KaiserNevinsPearlstein1983PF_26_351, %% quadrupole, FLR, m=1
    Berk+1985PPCNFR_2_321,% drift-modes, AIC, быстрые эл., Wall stabilization
    Berk+1984PF_27_2705, % wall stabilizatio, sharp boundary
    DIppolitoHafizi1981PF_24_2274, % sharp boundary, lom m, wall; не заметили эффект стеночной стабилизации (?)
    DIppolitoMyra1984PF_27_2256,% any m, sharp boundary hollow profile, wall stabilizxation, 2nd stability zone; RF stabilization
    KaiserPearlstein1985PhysFluids_28_1003,% m=1, eq. for arb. radial profile, FLR included; sharp boundary case;
    Kesner1985NF_25_275,% m=1, sharp boundary, parabolic axial profile of B; wall stabilization
    LiKesnerLane1985NF_25_907, % m=1, wall stabilization; CONDUCTING-WALL AND PRESSURE PROFILE EFFECT ON MHD STABILIZATION OF AXISYMMETRIC MIRROR
    LiKesnerLane1987NF_27_101,% partially enclosed wall. m=1; stepwise
    LiKesnerLoDestro1987NF_27_1259,% m=1, wall stabilization, sharp boundary;
    LoDestro1986PF_29_2329,% Вывод уравнения для жесткой моды
    ArseninKuyanov1996PPR_22_638%,% m=1 axisymmetric; ссылка на Лодестро
}.
%Kaiser, Nevins and Pearlstein investigated the stability of this mode in an open quadrupole trap in the low-pressure plasma approximation at $ \beta \to0 $ \cite{KaiserNevinsPearlstein1983PF_26_351}. In papers by
%     D'Ippolito and Myra \cite{DIppolitoMyra1984PF_27_2256},
%     Kaiser and Pearlstein \cite{KaiserPearlstein1985PhysFluids_28_1003},
%     Kesner \cite{Kesner1985NF_25_275},
%     Lee, Kesner and Lane \cite{LiKesnerLane1985NF_25_907},
%     Lee, Kesner and LoDestro \cite{LiKesnerLoDestro1987NF_27_1259}
%the wall stabilization of plasma with arbitrary and large beta is investigated.
%In the article by L.L.~Lodestro \cite{LoDestro1986PF_29_2329} a very nontrivial derivation of the equation for the rigid mode of ballooning oscillations in plasma with an arbitrary radial pressure profile is made.
%Arsenin and Kuyanov in their work \cite{ArseninKuyanov1996PPR_22_638} investigated the stability of the ballooning mode $ m = 1 $ in an axially symmetric trap taking into account the effects of finite beta.
%
These and some other papers on the stability of the $m=1$ ballooning mode and wall stabilization will be reviewed in a forthcoming paper.

%wall stabilization:
%\\
%D'Ippolito and Myra PF 1984 #9 \cite{DIppolitoMyra1984PF_27_2256} inverse radial (annular) profile and low‐m wall stabilization, arbitrary beta;
%\\
%Kaiser Pearlstein 1985 PF #3 \cite{KaiserPearlstein1985PhysFluids_28_1003}, arbitrary beta, wall stabilization, m=1;
%\\
%Кеснер 1985 NF #3 \cite{Kesner1985NF_25_275}, arbitrary beta, wall stabilization, m=1.
%\\
%Ли, Кеснера и Лэйна NF 1985 #8 \cite{LiKesnerLane1985NF_25_907},
%\\
%Li Kesner LoDestro 1987 NF #8 \cite{LiKesnerLoDestro1987NF_27_1259}, wall stabilization.

%В дальнейшем мы будем придерживаться следующего порядка изложения. Уравнение баллонных колебаний в пределе  плазмы низкого давления записано в разделе \ref{s2}. Его решения для некоторых модельных профилей магнитного поля и давления представлены в разделе \ref{s3}. Здесь вычислено критическое значение бета $\beta_{\text{crit}}$, выше которого баллонные возмущения становятся неустойчивыми. Оказалось, что вычисленные значения $ \beta_{\text {crit}} $ в некоторых случаях превышают единицу. Это означает, что предположение $\beta \ll 1$, сделанное при выводе упрощенного уравнения, не выполняется. Тем не менее приближение плазмы низкого давления дает количественно правильный результат для плазмы с радиальным профилем давления, близким к ступенчатому, для которой критическая бета мала.

In what follows, we will adhere to the following order of presentation. The equation for ballooning oscillations in the low-pressure plasma limit ($\beta \ll 1 $) is written in section \ref {s2}. Its solutions for some model magnetic field and pressure profiles are presented in section \ref {s3}. Here the critical value of beta $\beta_{\text{crit}} $ is calculated, above which ballooning disturbances become unstable. It turned out that the calculated values of $\beta_{\text{crit}} $ in some cases exceed one. This means that the assumption $\beta \ll 1 $, made when deriving the simplified equation, does not hold. Nevertheless, the low-pressure plasma approximation gives a quantitatively correct result for a plasma with a radial pressure profile close to a stepwise one, for which the critical beta is small.

%Уравнения в параксиальном приближении для мелкомасштабных баллонных возмущений в аксиально-симметричной плазме с конечным давлением и результаты, полученные в работе \cite{BushkovaMirnov1985BINP_103, BushkovaMirnov1986VANT_2_19} описаны в разделе \ref{s4}. Метод,

Equations in the paraxial approximation for small-scale ballooning perturbations in an axially symmetric plasma with finite pressure ($\beta \lesssim 1 $) and the results obtained in \cite{BushkovaMirnov1985BINP_103e, BushkovaMirnov1986VANT_2_19e}  for a model magnetic field are extended in Section~\ref{s4}. In Section~\ref{s5} critical beta is computed  for realistic magnetic field of the gas-dynamic trap. Main results are summarized in Section~\ref{s9}.

%В в разделе \ref{s5} воспроизведено выражение для потенциальной энергии желобковых и баллонных возмущений в непараксиальной аксиально-симметричной открытой ловушке в приближении бесконечного азимутального числа, полученное в классической работе \cite{Bernstein+1958ProcRSos_17_244} И.",Бернштейна с соавторами.

%В разделе \ref{s7} сделана оценка предельного бета в диамагнитной ловушке.
%
%In Section~\ref{s7}, an estimate of the critical beta in a diamagnetic trap proposed by Alexey Beklemishev in \cite{Beklemishev2016PoP_23_082506} is made.

%По словам Мирнова, возможно, что МГД-устойчивость непараксиальных систем рассматривали Арсенин и Сковорода.
%
%\begin{quote}
%  Hello Igor,
%
%Thank you for our valuable Skype talk. Attached please find one of Arsenin's paper about non-paraxial effects
%as well as Hasegawa's original paper on mirror instability. Because of grandchildren living with us almost 24/7 I had no time to start reviewing other non-paraxial papers. I will find  Callen-Hegna paper with coupling to sound waves. Also I will find my paper with Svidzinski where it was shown that for finite length pinch with end walls (non-periodic in z problem) div v is nonzero and contributes to marginal stability condition.
%
%All the best,
%Vladimir
%\end{quote}

%\section{Уравнение баллонных колебаний в плазме низкого давления}
\section{Equation of ballooning modes in low-pressure plasma}
\label{s2}

\begin{figure}
\includegraphics*[width=\linewidth,trim=0bp 0bp 0bp 40bp]{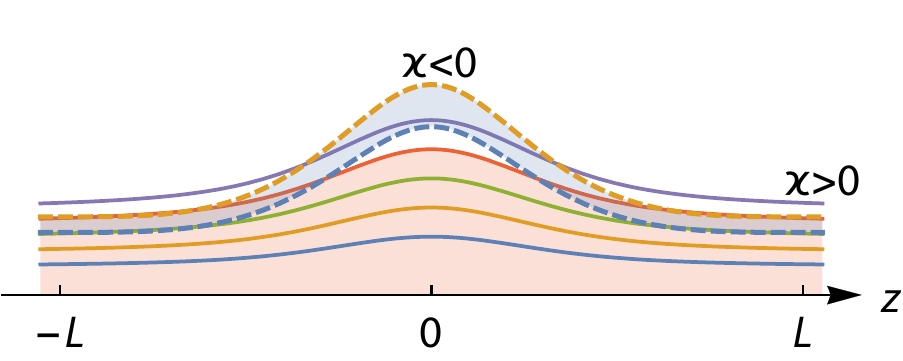}
    \caption{
        Ballooning perturbations distort the magnetic field and are localized in the region of unfavorable curvature, where $ \varkappa <0 $; the plasma tube, shown by the dotted line, floats up in the region of negative curvature away from the plasma axis, but is fixed at the ends of the trap.
    }
    \label{fig:214-01}
\end{figure}
In contrast to flute perturbations, ballooning modes are localized in the region of unfavorable curvature $\varkappa$ of magnetic field lines (where $ \varkappa <0 $), distorting the magnetic field approximately as shown in Fig.~\ref{fig:214-01}, approaching zero amplitude in the region of favorable curvature  (where $\varkappa > 0 $). Deformation of magnetic field lines requires energy expenditure, the source of which is the internal plasma energy $ p/(\gamma-1)$, where $p$ is the plasma pressure and $\gamma $ is the specific heat ratio; therefore ballooning-type disturbances are unstable only at a finite value of relative plasma pressure $\beta = 8\pi p/B^{2}$, exceeding some limit $ \beta_\text{crit} $:
    \begin{gather}
    \label{214.1:1}
    \beta > \beta_\text{crit}
    \qquad
    \text{(unstable)}
    .
    \end{gather}

The goal of exact theory is to calculate the $ \beta_\text{crit} $ parameter, but we will restrict ourselves to simple reasoning and order of magnitude estimates. In particular, for simplicity, we assume (as it was done in \cite{RyutovStupakov1980BINP_130e, RyutovStupakov1981IAEA_1_119} and many other papers) that ideally conducting ends are placed directly in the plugs in order to provide stability of the flute modes due to ``line-tying'' \cite{KunkelGuillory1965UCRL-16055, Ryutov+2011PoP_18_092301}. Thus, we narrow the class of possible perturbations allowed to compete while minimizing the potential energy of perturbations, so the found values of $ \beta_ \text{crit} $ will actually give an upper bound for this value. At the same time, the difference between the true values of $ \beta_\text{crit} $ from this upper boundary should be small, since behind the plug there is a deep (and plasma-filled) magnetic well (a region of favorable curvature), as a result of which the magnetic field lines are actually rigidly fixed in end plugs, returning us to the condition of being frozen into the ends.

In accordance with the real situation for devices of the GDT type, we will assume (following again Refs. \cite{RyutovStupakov1980BINP_130e, RyutovStupakov1981IAEA_1_119} and many subsequent papers) that the magnetic field lines make a small angle with the magnetic axis of the system (paraxial approximation) and that the distance between the magnetic plugs $ 2L $ is greater than the total plugs length $ 2L_{m} $, which can be separated by a quasi-uniform part. The small parameter characterizing the accuracy of the paraxial approximation is $ a/L_\text{m} $, where $ a $ is the transverse plasma size in the homogeneous part of the trap and $ L_\text{m} <L $ is the plug length. In the paraxial approximation, integration over the field line (over $ \dif {l} $) can be replaced by integration over the trap axis (over $ \dif {z} $). Let us agree that the coordinates $ z = \pm L $ correspond to the magnetic mirrors, and the homogeneous part of the trap occupies the region $ | z | <L-L_\text{m} $. Assuming also that $ \beta \ll 1 $ and the magnetic field even inside the plasma is close to vacuum field, we will assume that it is described with sufficient accuracy by the function $ B = B (z) $, which specifies the axial profile of the magnetic field on the trap axis. In this approximation, the reduced (that is, divided by $ 2 \pi $) magnetic flux at a distance $ r $ from the axis is equal to $ \Psi = r^{2} B (z)/2 $, and the field line equation
    \begin{equation}
    %\label{2:01}
    r(z) = r_{0}\,q(z),
    \qquad
    q(z)=\sqrt{B_{0}/B(z)}
    \end{equation}
%получается из условия постоянства магнитного потока внутри силовой трубки  $\Psi=r_{0}^{2}B_{0}/2=\const$, где $r_{0}$ и $B_{0}$ имеют смысл радиуса силовой линии и величины магнитного поля в медианной плоскости $z=0$.
%
is obtained from the condition of constancy of the magnetic flux inside the flux tube $ \Psi = r_{0}^{2} B_{0}/2 = \const $, where $ r_{0} $ and $ B_{0} $ have the meaning of the radius of the flux lines and values of the magnetic field in the median plane $ z = 0 $.

%Для возмущений желобкового типа в аксиально-симметричном пробкотроне для нормальной к силовой линии проекции вектора смещения $\vec{\xi}$ имеет место соотношение
%
For perturbations of the flute type in an axially symmetric mirror cell, for the projection of the displacement vector $ \vec {\xi} $ normal to the field line, the following relation holds:
    \begin{gather}
    %\label{2:02}
    \xi_{n}rB
    =
    \xi_{0}r_{0}B_{0}
    =
    \const
    .
    \end{gather}
%Отсюда в параксиальном приближении находим
Hence, in the paraxial approximation, we find
    \begin{gather}
    \label{2:03}
    \xi_{n}(z)
    =
    %\xi_{0}\sqrt{\frac{B_{0}}{B(z)}}
    \xi_{0}\,q(z)
    \equiv
    \xi_\text{fl}(z)
    ,
    \end{gather}
%где амплитуда смещения силовой трубки $\xi_{0}=\xi_{0}(\Psi,\theta)$ предполагается функцией магнитного потока $\Psi$ и азимутального угла $\theta $. В случае мелкомасштабного возмущения амплитуда $\xi_{0}$ резко пикирована (локализована) вблизи некоторых значений $\Psi_{\ast}$ и $\theta_{\ast}$, но до поры до времени этот факт не существен для последующих вычислений.
%
where the amplitude of the flux tube displacement $ \xi_{0} = \xi_{0} (\Psi, \theta) $ is assumed to be a function of the magnetic flux $ \Psi $ and the azimuthal angle $ \theta $. In the case of a small-scale perturbation, the $ \xi_{0} $ amplitude is sharply dived (localized) near some values of $ \Psi_{\ast} $ and $ \theta_{\ast} $, but for the time being this fact is not essential for subsequent calculations.

%Смещение \eqref{2:03} не удовлетворяет условию вмороженности $\xi_{n}(\pm L)=0$ на торцах, поэтому вместо \eqref{2:03} следует рассмотреть смещение более общего вида
%
The displacement \eqref{2:03} does not satisfy the condition of freezing $ \xi_{n} (\pm L) = 0 $ at the ends, therefore instead of \eqref{2:03} one should consider a displacement of a more general form
    \begin{gather}
    \label{2:04}
    \xi_{n}(z)=\alpha(z)\, \xi_\text{fl}(z)
    ,
    \end{gather}
%где $\alpha(z)$ "--- неизвестная пока безразмерная функция.  Наложим на неё два граничных условия:
where $ \alpha (z) $ is a still unknown dimensionless function. We impose two boundary conditions on it:
    \begin{gather}
    \label{2:05}
    \alpha(\pm L)=0,
    \qquad
    \alpha(0)=1
    .
    \end{gather}
%Первое из них обеспечивает выполнение условия вмороженности в торцы, а второе имеет смысл условия нормировки "---  оно фиксирует величину смещения в центре ловушки. Учитываю симметрию задачи, предположим, что искомая функция $\alpha (z)$ чётная, т.е.\ $\alpha (-z)=\alpha (z)$, поэтому достаточно найти решение лишь для одной половины ловушки, скажем, для $0<z<L$. Нечётное решение $\alpha (-z)=-\alpha (z)$ также существует. Оно получается для граничных условий $\alpha(\pm L)=\alpha(0)=0$, однако ему соответствует б\'{о}льшее по величине значение $\beta_{\text{crit}}$.
%
The first of them ensures the fulfillment of the condition of being frozen into the ends, and the second has the meaning of the normalization condition, it fixes the displacement at the center of the trap. Taking into account the symmetry of the problem, we assume that the sought function $ \alpha (z) $ is even, that is, $ \alpha (-z) = \alpha (z) $, so it suffices to find a solution for only one half of the trap, say, for $ 0 <z <L $. An odd solution $ \alpha (-z) = - \alpha ( z) $ also exists. It is obtained for the boundary conditions $ \alpha (\pm L) = \alpha (0) = 0 $, but it corresponds to a larger value of $ \beta_{\text{crit}}$.

%Как показано в \cite[\S29]{Kotelnikov2020V2} (см. также \cite{RyutovStupakov1980BINP_130, RyutovStupakov1981IAEA_1_119}), энергия возмущения баллонного типа равна
As shown in \cite [\S29] {Kotelnikov2020V2e} (see also \cite{RyutovStupakov1980BINP_130e, RyutovStupakov1981IAEA_1_119}), the energy of the ballooning-type perturbation is
    \begin{gather}
    %\label{214.2:16'}
    \label{2:06}
    {W}_{F} =
    %{W}_{B}+{W}_{P}  =
    \left(
        \frac{1}{8\pi}
        \iint \dif{\theta } \dif{\Psi}\,
        \xi_{0}^{2}B_{0}
    \right)
    \left(
        \int \dif{z}
        \left\{
            \left( \alpha' \right)^{2}
            +
            2\overline\beta
            q^{3} q''
            \alpha^{2}
        \right\}
    \right)
    ,
    \end{gather}
%где
where
\begin{gather}
    \label{2:07}
    \overline\beta
    =
    \iint \dif{\theta } \dif{\Psi}\,\xi_{0}^{2}\, % B_{0}
    \beta(\Psi)
    \Bigm/
    \iint \dif{\theta } \dif{\Psi}\,\xi_{0}^{2} % B_{0}
    ,\\
    \label{2:08}
    \beta(\Psi)
    =
    - \frac{8\pi \Psi}{B_{0}^{2}}\,\der{p}{\Psi}
%    =
%    \beta_{0}\,\frac{\Psi}{p_{0}}\,\left|\der{p}{\Psi}\right|
%    ,\\
%    \beta_{0} = \frac{8\pi p_{0}}{B_{0}^{2}}
    ,
    \end{gather}
%а штрих обозначает производную по $z$.
and the prime denotes the derivative with respect to $ z $.

%В соответствии с энергетическим принципом \cite{Bernstein+1958ProcRSos_17_244}, система устойчива, если минимальное значение интеграла \eqref{2:06} больше нуля. Множитель внутри первой пары скобок в \eqref{2:06} заведомо больше нуля, поэтому знак $W_{F}$ определяется знаком интеграла внутри второй пары скобок.  Чтобы найти минимум $W_{F}$, возьмём вариацию интеграла
%
According to the energy principle \cite{Bernstein+1958ProcRSos_17_244}, the system is stable if the minimum value of the integral \eqref{2:06} is greater than zero. The factor inside the first pair of parentheses in \eqref{2:06} is certainly greater than zero, so the sign of $ W_{F} $ is determined by the sign of the integral inside the second pair of parentheses. To find the minimum of $ W_{F} $, we take the first variation of the integral \eqref{2:06},
    \begin{multline*}
    %\label{2:09}
    \delta
    \int_{0}^{L}
    \dif{z}
    \left\{
        \left(
            \alpha'
        \right)^{2}
        +
        2\overline\beta
        \alpha^{2}
        q^{3} q''
    \right\}
    =\\=
    2
    \int_{0}^{L}
    \dif{z}
    \left\{
        \alpha'\,
        \delta\alpha'
        +
        2\overline{\beta}
        q^{3}q''
        \alpha\,\delta\alpha
    \right\}
    ,
    \end{multline*}
%и приравняем её к нулю. Так мы получим уравнение, которому должна удовлетворять экстремаль, т.",e.\ функция $\alpha(z)$, доставляющая минимум энергии возмущения ${W}_{F}$. Поскольку значения функции $\alpha(z)$ на концах интервала интегрирования заданы граничными условиями \eqref{2:05}, вычисляя вариацию, нужно считать, что $\delta\alpha(0)=\delta\alpha(L)=0$. Выполнив с учётом этого факта  интегрирование по частям, получим интеграл
%
and set it to zero. In this way we obtain an equation that the extremal must satisfy, that is, the function $ \alpha (z) $, which minimizes the perturbation energy $ {W}_{F} $. Since the values of the function $ \alpha (z) $ at the ends of the integration interval are given by the boundary conditions \eqref{2:05}, calculating the variation, we must assume that $ \delta \alpha (0) = \delta \alpha (L) = 0 $. Taking this fact into account and integrating by parts, we get the integral
    \begin{gather*}
    %\label{2:11}
    2
    \int_{0}^{L}
    \dif{z}
    \left\{
        -
        \alpha''
        +
        2\overline{\beta}
        q^{3}q''
        \alpha
    \right\}
    \delta \alpha
    ,
    \end{gather*}
%который обращается в ноль при любой вариации $\delta\alpha$, если
which vanishes for any variation of $ \delta\alpha $ if
    \begin{gather}
    \label{2:12}
    \alpha''
    -
    2\overline{\beta}
    q^{3} q''\,
    \alpha
    =
    0
    .
    \end{gather}
%Как видно, вычисление предельного значения $\overline{\beta}_{\text{crit}}$ параметра $\overline{\beta}$, при котором существует отличное от нуля решение уравнения \eqref{2:12} с указанными выше граничными условиями, сводится к квантовомеханической задаче об определении условий возникновения уровня нулевой энергии в потенциале
%
As one can see, the calculation of the critical value $ \overline {\beta}_{\text{crit}} $ of the parameter $ \overline {\beta} $, for which there is a nonzero solution to Eq.~\eqref{2:12} with the above boundary conditions \eqref{2:05}, is reduced to the quantum mechanical problem of determining the conditions for the occurrence of a zero energy level in the potential
    \begin{equation}
    \label{2:14}
    V(z) =
    \begin{cases}
      \infty        , & \mbox{if } z<0 ;\\
      2\overline{\beta} q^{3}q'', & \mbox{if } 0 < z < L ;\\
      \infty        , & \mbox{if } z>L
      .
    \end{cases}
    \end{equation}
%Можно также сказать, что $\overline{\beta}_{\text{crit}}$ является собственным значением классической задачи Штурма-Лиувилля для уравнения \eqref{2:12} с граничными условиями \eqref{2:05}, а решение этой задачи $\alpha (z)$ является собственной функцией. Нетрудно доказать, что подстановка собственной функции  и собственного значения в интеграл \eqref{2:06} обращает в ноль энергию баллонного возмущения, что в согласии с энергетическим принципом соответствует пограничному состоянию между устойчивостью и неустойчивостью.
%
We can also say that $ \overline {\beta}_{\text{crit}} $ is the eigenvalue of the classical Sturm-Liouville problem for Eq.~\eqref{2:12} with boundary conditions \eqref{2:05}, and the solution to this problem $ \alpha (z) $ is an eigenfunction. It is easy to prove that the substitution of the eigenfunction and the eigenvalue in the integral \eqref{2:06} makes the energy of the ballooning disturbance to zero, which, in accordance with the energy principle, corresponds to the marginal state between stability and instability.

%Предельное значение $\overline{\beta}_{\text{crit}}$ вычислено в разделе \ref{s3}, а фактическая величина параметра $\overline{\beta }$, которую нужно сравнивать с предельным значением, зависит от вида функции $\xi_{0}=\xi_{0}(\Psi,\theta )$ и профиля давления $p(\Psi)$.

The value of $ \overline {\beta}_{\text{crit}} $  is calculated in Section~\ref{s3}, and the actual value of the $ \overline {\beta} $ parameter to be compared with the critical value depends on the spacial profile  of function $ \xi_{0} = \xi_{0} (\Psi, \theta) $ and pressure profile $ p (\Psi) $.

%\section{Предельное \texorpdfstring{$\beta$}{бета} в плазме низкого давления}
\section{critical beta in low pressure plasma}
%\section{Low-pressure plasma approximation}
\label{s3}

%В наше время методы компьютерных вычислений достигли такой стадии совершенства, что любой человек, не владеющий продвинутыми методами решения дифференциальных уравнений, может почувствовать себя великим математиком, написав за полчаса программу для вычисления собственных функций и собственных значений задачи Штурма-Лиувилля. Тем не менее мы воспользуемся представившимся случаем, чтобы наглядно продемонстрировать метод приближённого вычисления $\overline{\beta}_{\text{crit}}$ с помощью вариационного принципа.

%Nowadays, computer computing methods have reached such a stage of perfection that anyone who does not know advanced methods for solving differential equations can feel like a great mathematician, having written a program for calculating eigenfunctions and eigenvalues of the Sturm-Liouville problem in half an hour. Nevertheless, we will use the presented case to illustrate the method of approximate calculation of $ \overline {\beta}_{\text{crit}} $ using the variational principle.

\begin{figure}
    \includegraphics[width=\linewidth]{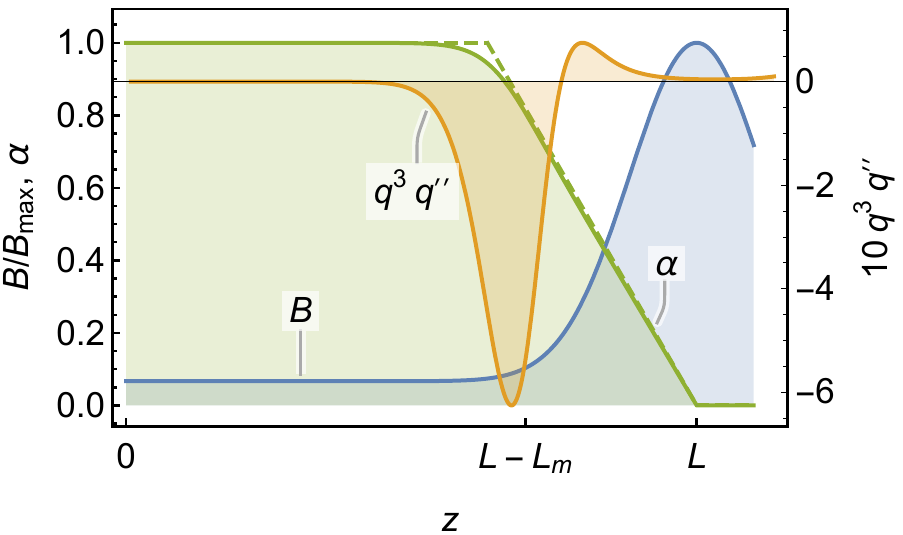}
    \caption{
        %Магнитное поле \eqref{3:01}, эффективный потенциал $q^{3}q''$ и собственная функция $\alpha(z)$ в задаче Штурма-Лиувилля для случая $K=15$, $L=6$, $\Delta L=1$; пунктиром показан график функции \eqref{3:02} при $L_{m} = 2.2\Delta L$.
        %
        Magnetic field \eqref{3:01}, effective potential $ q^{3} q '' $ and eigenfunction $ \alpha (z) $ in the Sturm-Liouville problem for the case $ K = 15 $, $ L = 6 $ , $ \Delta L = 1 $; the dotted line shows the graph of the function \eqref{3:02} for $ L_{m} = 2.2 \Delta L $.
    }\label{fig:214-02-03}
\end{figure}
%Для первого примера выберем зависимость магнитного поля от координаты $z$ на оси ловушки в виде
For the first example, let us choose the dependence of the magnetic field on the coordinate $ z $ on the trap axis in the form
    \begin{equation}
    \label{3:01}
    B(z)/B_{0}
    =
    1+(K-1) \left(
        \e^{-(z-L)^2/\Delta L^{2}}+\e^{-(L+z)^2/\Delta L^{2}}
    \right),
    \end{equation}
%где $K$ имеет смысл пробочного отношения, а $2L$ есть расстояние между магнитными пробками. Для такого поля функция эффективного потенциала $q^{3}q''$ приблизительно равна нулю вне пробки (при $0<z<L-L_{\text{m}}$) и быстро спадает вглубь пробки вместе с ростом пробочного отношения $B/B_{0}=1/q^{2}$, как показано на рис.~\ref{fig:214-02-03}. На входе в пробку на графике $q^{3}q''$ имеется яма. Её ширина  приблизительно равна $L_{\text{m}}$, а сама яма расположена в области пробки, примыкающей к однородному магнитному полю, там, где $B/B_{0}\sim 1$ и $z\approx L-L_{\text{m}}$. Вне области потенциальной ямы уравнение \eqref{3:03} сводится к тривиальному равенству $\alpha ''=0$, поэтому там $\alpha (z)$ приблизительно будет линейной функцией координаты $z$. С двух сторон от потенциальной ямы производная $\alpha '$ является константой, но константа меняется в области ямы. Сконструируем из линейных функций тестовую функцию
%
where $ K $ has the meaning of the mirror ratio, and $ 2L $ is the distance between the magnetic mirrors. For such a field, the effective potential function $ q^{3} q '' $ is approximately equal to zero outside the plug (for $ 0 < |z| <L-L_{\text{m}} $) and rapidly decreases deep into the plug along with the growth of the mirror ratio $ B/B_{0} = 1/q^{2} $, as shown in Figure~\ref{fig:214-02-03}. There is a hole at the entrance to the magnetic mirror on the $ q^{3} q '' $ graph. Its width is approximately equal to $ L_{\text{m}} $, and the well itself is located in the region of the plug, adjacent to the uniform magnetic field, where $ B/B_{0} \sim 1 $ and $ z \approx L- L_{\text{m}} $. Outside the region of the potential well, the equation \eqref{3:03} reduces to the trivial equality $ \alpha '' = 0 $, so there $ \alpha (z) $ will be approximately a linear function of the coordinate $ z $. On both sides of the potential well, the derivative $ \alpha '$ is a constant, but the constant changes in the region of the well. Let us construct a test function from linear functions
    \begin{equation}
    \label{3:02}
    \alpha_{1}(z) =
    \begin{cases}
        1 & |z| \leq L-L_m, \\
        \frac{L-\left|z\right|}{L_m} & L-L_m<\left| z\right| \leq L,
    \end{cases}
    \end{equation}
%обладающую указанными свойствами, подставим её в интеграл \eqref{2:06} и приравняем результат к нулю. Из полученного уравнения находим приближённое значение предельного бета:
possessing the indicated properties, we substitute it into the integral \eqref{2:06} and equate the result to zero. From the resulting equation, we find the approximate value of the critical beta:
    \begin{equation}
    \label{3:03}
    \overline{\beta}_{\text{crit}}
    =
    -
    \frac{1}{2}
    {
        \int_{0}^{L}
        \dif{z}
        \left(
            \alpha_{1}'
        \right)^{2}
    }
    \Bigm/
    {
        \int_{0}^{L}
        \dif{z}\,
        \alpha_{1}^{2}
        q^{3} q''
    }
    .
    \end{equation}
%Для магнитного поля \eqref{3:01} вычисленная таким образом величина $\overline{\beta}_{\text{crit}} = \num{0.5795}$ лишь незначительно отличается от точного собственного значения $\overline{\beta}_{\text{crit}} = \num{0.597052}$. График собственной функции $\alpha(z)$ показан на рис.~\ref{fig:214-02-03}, там же пунктиром показана тестовая функция $\alpha_{1}(z)$.
%
For the magnetic field \eqref{3:01}, the calculated value $ \overline {\beta}_{\text{crit}} = \num {0.5795} $ only slightly differs from the exact eigenvalue $ \overline {\beta}_{\text{crit}} = \num {0.597052} $. The graph of the eigenfunction $ \alpha (z) $ is shown in Fig.~\ref{fig:214-02-03}, where the dashed line shows the test function $ \alpha_{1} (z) $.

%\begin{figure}
%    \includegraphics[width=\linewidth]{2020-214-03}
%    \caption{
%        График собственной функции $\alpha(z)$ задачи Штурма-Лиувилля для профиля магнитного поля, который изображён на рис.~\ref{fig:214-02}. Пунктиром показан график тестовой функции \eqref{3:05}.
%    }\label{fig:214-03}
%\end{figure}

\begin{figure}
    \includegraphics[width=\linewidth]{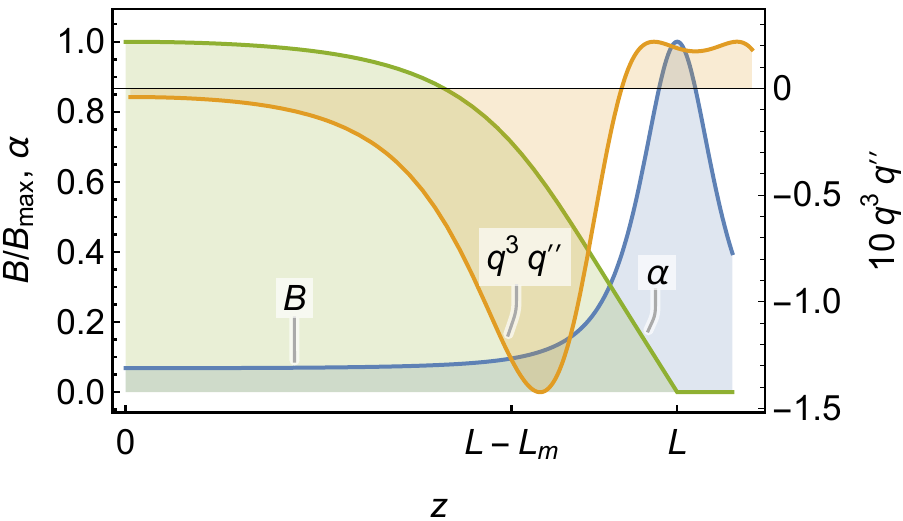}
    \caption{
        %Пробкотрон "--- система двух соосных круглых катушек.
        %        Магнитное поле \eqref{3:04}, эффективный потенциал $q^{3}q''$ и собственная функция $\alpha(z)$ в задаче Штурма-Лиувилля для случая  $K=15$, $L=6$, $b=0.6$, $L_{m}\approx 3b$.
        %
        Mirror cell is a system of two coaxial round coils.
        Magnetic field \eqref{3:04}, effective potential $ q^{3} q '' $ and eigenfunction $ \alpha (z) $ in the Sturm-Liouville problem for the case $ K = 15 $, $ L = 6 $ , $ b = 0.6 $, $ L_{m} \approx 3b $.
    }\label{fig:214-04-05}
\end{figure}
%Расчёты показывают, что предельное бета весьма чувствительно к профилю магнитного поля. Этот факт подтверждает второй пример с магнитным полем
Calculations show that the critical beta is very sensitive to the magnetic field profile. This fact is confirmed by the second example with a magnetic field
\begin{equation}
    \label{3:04}
    \frac{B(z)}{B_{0}} =
    1
    +
    (K-1)
    \left[
        \frac{b^3}{\left(b^2+(z-L)^2\right)^{3/2}}
        +
        \frac{b^3}{\left(b^2+(L+z)^2\right)^{3/2}}
    \right]
    ,
\end{equation}
%которое моделирует пробкотрон, составленный двумя катушками. В этом примере предельное значение $\overline{\beta}_{\text{crit}}=\num{1.17645}$ формально превышает единицу.
which simulates a mirror cell composed of two coils. In this example, the critical value $ \overline {\beta}_{\text{crit}} = \num {1.17645} $ is formally greater than one.

%Обсудим полученные результаты. Вычисленное значение $\overline{\beta}_{\text{crit}}$ оказалось близко к единице или даже больше единицы тогда как, приступая к выводу уравнения баллонных колебаний в разделе \ref{s2}, мы использовали приближение $\beta \ll 1$ и считали магнитное поле вакуумным. Следовательно, эта наша гипотеза была не совсем верна. Тем не менее наш труд не был совсем уж бесполезен.

Let us discuss the results obtained. The calculated value of $ \overline {\beta}_{\text{crit}} $ turned out to be close to one or even more than one, whereas, starting to derive the equation of ballooning oscillations in Section~\ref{s2}, we used the approximation $ \beta \ll 1 $ and assumed the magnetic field to be vacuum. Therefore, this hypothesis of ours was not entirely correct. Nevertheless, our work was not entirely useless.

%Во-первых заметим, что параметр $\overline{\beta}$, как и функция $\beta(\Psi)$, которые определены формулами \eqref{2:07} и \eqref{2:08}, не равны параметру бета в традиционном понимании
%
First, note that the $ \overline {\beta} $ parameter, like the $ \beta (\Psi) $ function, which are defined by the formulas \eqref{2:07} and \eqref{2:08}, are not equal to the parameter beta in the traditional sense
    \begin{equation}
    \label{3:05}
    \beta_{0}=\frac{8\pi p_{0}}{B_{0}^{2}}
    ,
    \end{equation}
%где $p_{0}$ имеет смысл давления плазмы на её оси при $\Psi=0$. В частном случае, когда давление имеет степенной профиль
%
where $ p_{0} $ has the meaning of the plasma pressure on its axis at $ \Psi = 0 $. In the particular case when the pressure has a power-law profile
    \begin{equation}
    \label{3:06}
    p(\Psi)=p_{0}\,(1-\Psi^{k}/\Psi_{a}^{k})
    \end{equation}
%при $\Psi<\Psi_{a}$, максимальное значение  $\beta(\Psi)$ достигается на крайней силовой линии $\Psi=\Psi_{a}$, где $\beta(\Psi_{a})=k\beta_{0}$. Если $k=1$ (параболический по $r$ профиль давления) имеет место равенство $\beta(\Psi_{a})=\beta_{0}$, но для более крутого профиля с $k>1$ величина  $\beta(\Psi_{a})$ превышает $\beta_{0}$. Точно так же для крутого профиля давления, близкого к ступеньке, величина $\overline{\beta}$ может быть значительно больше $\beta_{0}$, поэтому возможна ситуация, когда $\overline{\beta}\sim1$ при том, что $\beta_{0}\ll1$ и приближением вакуумного магнитного поля вполне можно пользоваться.
%
for $ \Psi <\Psi_{a} $, and $p(\Psi)=0$ for $ \Psi >\Psi_{a} $, the maximum value of $ \beta (\Psi) $ is reached on the boundary field line $ \Psi = \Psi_{a} $, where $ \beta (\Psi_{a}) = k \beta_{0} $. If $ k = 1 $ (pressure profile parabolic in $ r $), the equality $ \beta (\Psi_{a}) = \beta_{0} $ holds, but for a steeper profile with $ k> 1 $ the value $ \beta (\Psi_{a}) $ is greater than $ \beta_{0} $. In the same way, for a steep pressure profile close to a step, the value of $ \overline {\beta} $ can be much larger than $ \beta_{0} $, so a situation is possible where $ \overline {\beta} \sim 1 $ whereas $ \beta_{0} \ll1 $ and the vacuum magnetic field approximation can be completely valid.

%Ступенчатый профиль давления можно рассматривать как предел близкого к ступеньке трапециевидного распределения с малой, но конечной шириной граничного слоя $\Delta\Psi$. Для определённости примем, что давление плазмы равно $p_{0}$ при $\Psi<\Psi_{a}$ и линейно уменьшается до нуля в пограничном слое с шириной $\Delta \Psi$. Результат вычисления $\overline{\beta}$ зависит от соотношения $\Delta\Psi$ и размера желобка $\delta\Psi$ по переменной $\Psi$.

The stepped pressure profile can be considered as the limit of a trapezoidal distribution close to a step with a small but finite width of the boundary layer $ \Delta \Psi $. For definiteness, let us assume that the plasma pressure is equal to $ p_ {0} $ at $ \Psi < \Psi_{a} $ and decreases linearly to zero in the boundary layer with the width $ \Delta\Psi $. The result of calculating $ \overline {\beta} $ depends on the ratio of $ \Delta \Psi $ and the size $\delta \Psi $ of the perturbation with respect to the variable $ \Psi $.

%Если размер желобка мал по сравнению с шириной границы, $\delta\Psi\ll\Delta \Psi$, локальный параметр \eqref{2:08} в интеграле \eqref{2:07} можно считать неизменным на размере желобка, так что $\overline{\beta}\approx\beta(\Psi_{\ast})$, где $\Psi_{\ast}$ "--- значение $\Psi$ на силовой линии, вблизи которой локализовано возмущение. Локальное значение
%
If the size of the flute tube is small compared to the width of the border, $ \delta \Psi \ll \Delta \Psi $, the local parameter \eqref{2:08} in the integral \eqref{2:07} can be considered constant across the size of the tube, so that $ \overline {\beta} \approx \beta (\Psi_{\ast}) $, where $ \Psi_{\ast} $ is the value of $ \Psi $ on the field line, near which the disturbance is localized. Local value
    \begin{equation}
    %\label{3:11}
    \beta(\Psi_{\ast}) \approx
    %\frac{8\pi p_{0}}{B_{0}^{2}}\,
    \beta_{0}\,
    \frac{\Psi_{a}}{\Delta\Psi}
    \end{equation}
%в пограничном слое на силовой линии, где локализовано возмущение, при малой ширине границы велико и может легко превысить критическое значение. Следовательно, мелкомасштабные возмущения будут неустойчивы, даже если давление плазмы существенно меньше давления магнитного поля, т.е.\ $\beta_{0}\ll1$.
%
in the boundary layer on the field line, where the disturbance is localized, is large for a small boundary width and can easily exceed the critical value. Consequently, small-scale perturbations will be unstable even if the plasma pressure is much lower than the magnetic field pressure, ie, $ \beta_{0} \ll1 $.

%Однако мелкомасштабные колебания вряд ли могут нанести большой ущерб равновесной конфигурации плазмы. Более опасны крупномасштабные возмущения. Если $\delta \Psi\gg\Delta \Psi$, выполняя интегрирование по $\Psi$ в формуле  \eqref{2:07}, можно считать, что
%
However, small-scale oscillations are unlikely to cause great damage to the equilibrium plasma configuration. Large-scale disturbances are more dangerous. If $ \delta \Psi \gg \Delta \Psi $, performing integration over $ \Psi $ in the formula \eqref{2:07}, we can assume that
    \begin{equation}
    \label{3:12}
    \beta(\Psi) =
    %\frac{8\pi p_{0}}{B_{0}^{2}}\,
    \beta_{0}\,
    \Psi_{a}\delta(\Psi-\Psi_{a})
    .
    \end{equation}
%Подстановка \eqref{3:12} в формулу \eqref{2:07} приводит к оценке
Substitution of \eqref{3:12} into the formula \eqref{2:07} results in the estimation
    \begin{equation}
    \label{3:14}
    \overline\beta
    \approx
    %\frac{8\pi p_{0}}{B_{0}^{2}}\,
    \beta_{0}\,
    \frac{\Psi_{a}}{\delta\Psi}
    .
    \end{equation}
%Из неё видно, что $\overline\beta$ уменьшается с увеличением масштаба возмущения, поэтому самые крупномасштабные возмущения при заданном значении $\beta_{0}$ могут быть устойчивы, тогда как мелкомасштабные возмущение будут неустойчивы. Таким образом, результатом развития баллонной неустойчивости будет установление «гладкого» радиального профиля давления.
%
It can be seen from it that $ \overline \beta $ decreases with an increase in the scale of the disturbance; therefore, the largest-scale perturbations for a given value of $ \beta_{0} $ can be stable, while small-scale disturbances will be unstable. Thus, the development of ballooning instability will result in the establishment of a “smooth” radial pressure profile.

%Есть и второй вывод, который можно сделать из установленного нами факта, что в приближении вакуумного магнитного поля формально $\overline{\beta}_{\text{crit}}\gtrsim 1$. Теперь мы можем утверждать, что предельное $\beta_{0}$ в аксиально-симметричном пробкотроне с гладким профилем давления не мало по сравнению с единицей. Чтобы правильно вычислить предельное $\beta_{0}$ необходимо отказаться от приближения вакуумного поля, которое также называют приближением плазмы низкого давления. Необходимые расчёты сравнительно просто удаётся выполнить для параксиальных открытых ловушек. Они описаны в следующем разделе.

There is also a second conclusion, which can be drawn from the fact established in the above, that in the approximation of a vacuum magnetic field it is formally $ \overline {\beta}_{\text{crit}} \gtrsim 1 $. Now we can assert that the critical $ \beta_{0} $ in an axially symmetric mirror cell with a smooth pressure profile is not small in comparison with unity. To correctly calculate the critical $ \beta_{0} $ it is necessary to abandon the vacuum field approximation, which is also called the low-pressure approximation. The necessary calculations are relatively easy to perform for paraxial mirror traps. These are described in the next section.

%\section{Предельное \texorpdfstring{$\beta$}{} в параксиальной открытой ловушке}
%\section{critical \texorpdfstring{$\beta$}{beta} in a paraxial miiror trap}
\section{Paraxial mirror trap}
\label{s4}

%В параксиальном приближении потенциальная энергия баллонных возмущений найдена в статье У.",Ньюкомба \cite{Newcomb1981JPP_26_529}. Для аксиально симметричной открытой ловушки с анизотропной плазмой Ньюкомб приводит следующее выражение (его формула 177)
%
In the paraxial approximation, the potential energy of ballooning perturbations is found in the article by William Newcomb \cite{Newcomb1981JPP_26_529}. For an axially symmetric open trap with anisotropic plasma, Newcomb gives the following expression (his formula 177)
    \begin{gather}
    \label{4:01}
    W_{F} =
    \frac{1}{2}
    \iiint \dif{\theta }
    \dif{\Psi}
    \dif{z}
    \left[
        \frac{Q}{B}
        \left(
            \frac{X'^{2}}{r^{2}B^{2}}
            +
            r^{2}Y'^{2}
        \right)
        -
        \frac{2\varkappa}{rB^{2}}
        \parder{\overline{p}}{\Psi}
        X^{2}
    \right]
    .
    \end{gather}
%Здесь предполагается,  что все функции кроме $B_{\text{vac}}(z)$ зависят от $\Psi$ и $z$, штрих обозначает частную производную по $z$ при фиксированной величине $\Psi$,
%
It is assumed here that the prime denotes the partial derivative with respect to $ z $, and all functions except for $B_{\text{vac}}=B_{\text{vac}}(z)$, $X=X(\theta ,\Psi,z)$,  and $Y=Y(\theta ,\Psi,z)$, depend on $ \Psi $ and $ z $,
    \begin{subequations}
    \begin{gather}
    \label{4:02}
    r^{2} = 2 \int_{0}^{\Psi} \frac{\dif{\Psi}}{B(\Psi,z)}
    ,\qquad
    B^{2} = B_\text{vac}^{2}(z) - 8\pi p_{\bot}
    ,\\
    \label{4:03}
    Q = B^{2} + 4\pi\left( p_{\bot} - p_{\|} \right)
    ,\qquad
    \overline{p} = \frac{p_{\bot}+p_{\|}}{2}
    ,\\
    \label{4:04}
    X = \vec{\xi}\cdot \nabla\Psi = rB\xi_{n},
    \qquad
    Y = \vec{\xi}\cdot \nabla\theta = \frac{\xi_{\theta }}{r}
    ,\\
    \label{4:05}
    \varkappa = r''
    .
    \end{gather}
    \end{subequations}
%Ньюкомб указывал, что функция $Q$ положительна при выполнении условия стабилизации шланговой неустойчивости
%
Newcomb pointed out that the function $ Q $ is positive under the condition of stabilization of the hose instability
    \begin{equation}
    \label{4:09}
    \beta_{\|} > 2 + \beta_{\perp}
    ,
    \end{equation}
%которое всегда предполагалось выполненным. В противном случае первое слагаемое в квадратных скобках в подынтегральном выражении \eqref{4:01} становится отрицательным (дестабилизирующим), как и второе слагаемое.
%
which was always assumed to be fulfilled. Otherwise, the first term in square brackets in the integrand \eqref{4:01} becomes negative (destabilizing), like the second term.

%Минимизация $W_{F}$ по $Y$ даёт $Y''=0$, что позволяет считать нулём слагаемое с $Y'^{2}$ в \eqref{4:01}. Расчёты для этого случая выполнены О.А.",Бушковой и В.В.",Мирновым в работе \cite{BushkovaMirnov1985BINP_103, BushkovaMirnov1986VANT_2_19}. Они исследовали предел изотропной плазмы, когда $p_{\bot}=p_{\|}=p(\psi)$ и переписали \eqref{4:01} в безразмерном виде, удобном для численных расчётов:
%
Minimizing $ W_{F} $ over $ Y $ gives $ Y'' = 0 $, which allows us to consider the term with $ Y '^{2} $ in \eqref{4:01} as zero. The calculations for this case were performed by Bushkova and Mirnov  \cite{BushkovaMirnov1985BINP_103e, BushkovaMirnov1986VANT_2_19e}. They investigated the limit of isotropic plasma, when $ p_{\bot} = p_{\|} = p (\psi) $ and rewrote Eq.~\eqref{4:01} in the dimensionless form convenient for numerical calculations:
    \begin{gather}
    \label{4:11}
    W_{F} =
    \frac{\Delta\psi}{8\pi}
    \int\dif{z}
    \left[
        \frac{X'^{2}}{r^{2}b}
        -
        \beta_{0}
        \frac{r''}{rb^{2}}
        \der{f}{\psi}
        X^{2}
    \right]
    ,
    \end{gather}
where $\psi=\Psi/\Psi_{a}$,
    \begin{subequations}
    \label{4:12}
    \begin{gather}
    \label{4:12a}
    \beta_{0} = {8\pi p_{0}}/{B_{\text{vac}0}^{2}}
    ,\qquad
    %\label{4:12b}
    f(\psi) = {p(\psi)}/{p_{0}}
    %= 1 - \frac{\Psi^{2}}{\Psi_{a}^{2}}
    ,\\
    \label{4:12c}
    b=B/B_{\text{vac}0}
    ,\qquad
    %\label{4:12d}
    b_{\text{vac}}=B_{\text{vac}}/B_{\text{vac}0}
    ,\\
    \label{4:12e}
    b(\psi,z) = \sqrt{b_\text{vac}^{2}(z)-\beta_{0} f(\psi)}
    ,\\
    \label{4:12f}
    r^{2}(\psi,z) = 2 \int_{0}^{\psi}
    \frac{\dif{\psi'}}{\sqrt{b_\text{vac}^{2}(z)-\beta_{0} f(\psi')}}
    .
    \end{gather}
    \end{subequations}
%Варьирование интеграла \eqref{4:11} приводит к уравнению
Varying the integral \eqref{4:11} leads to the equation
    \begin{equation}
    \label{4:14}
    \der{}{z}
    \left(
        \frac{1}{r^{2}b}
        \der{X}{z}
    \right)
    +
    \beta_{0}
    \frac{r''}{rb^{2}}
    \der{f}{\psi}
    X
    =0
    .
    \end{equation}
The same equation can be deduced from Eq.~(18) in Ref.~\cite{DIppolitoHafiziMyra1982PF_25_2223} if we put zero frequency, $\omega =0$. Calculations were performed for the pressure profile
    \begin{equation}
    \label{4:15}
    f(\psi) = 1 - \psi^{2}
    ,
    \end{equation}
which allows one to calculate the integral \eqref{4:12f} and write the equation of the field line:
    \begin{equation}
    \label{4:16}
    r^{2}(\psi,z)
    =
    \frac{2}{\sqrt{\beta_{0} }}
        \log
        \frac{
            \sqrt{
                b_{\text{vac}}(z){}^2
                -
                \beta_{0} \left( 1 - \psi^2\right)
            }
            +
            \sqrt{\beta_{0} }\, \psi
        }{
            \sqrt{b_{\text{vac}}^2(z)-\beta_{0} }
        }
       .
    \end{equation}
%Предельное бета было вычислено для крайней силовой линии при $\psi=0.99$ для семейства аксиальных профилей вакуумного магнитного поля
%
The critical beta was calculated for the boundary field line at $ \psi = 0.99 $ for a family of axial profiles of the vacuum magnetic field
    \begin{equation}
    \label{4:17}
    b_{\text{vac}}(z)
    =
    \left[
        1
        -
        \left(
            1 - K^{-\gamma/2}
        \right)
        z^{\mu}
    \right]^{-2/\gamma }
    \end{equation}
%с разными показателями степеней $\mu $ и $\gamma $. Результаты сведены в таблицу \ref{tab:beta}.
%
with mirror ratio $K=100$ and a set of parameters $ \mu $ and $ \gamma $. In order to check our code, we recalculate their results assuming the boundary conditions
    \begin{gather}
    \label{4:18}
    X(0)=1, \qquad X(1)=0
    \end{gather}
and obtained similar results which are listed in Table \ref{tab:beta2}.
%%%\begin{table}[!t]\medskip
%%%  \centering
%%%  \newcolumntype{Y}{>{\centering\arraybackslash}X}
%%%  \begin{tabularx}{\linewidth}{|c|Y|Y|Y|Y|Y|Y|}
%%%    \hline
%%%    $\gamma \setminus \mu$ & 1 & 2 & 3 & 4 & 5 & 6 \\ \hline
%%%    0.5 & 0.19 & 0.32 & 0.38 & 0.41 & 0.42 & 0.43 \\ \hline
%%%    2   & 0.36 & 0.49 & 0.53 & 0.55 & 0.56 & 0.56 \\ \hline
%%%    6   & 0.62 & 0.70 & 0.72 & 0.73 & 0.74 & 0.74 \\ \hline
%%%  \end{tabularx}
%%%  \caption{
%%%    %% Эта данные из \cite{BushkovaMirnov1985BINP_103e}
%%%    %Предельное бета для локальных возмущений на крайней силовой линии при различных значениях параметров $\mu$ и $\gamma $.
%%%    critical beta for local perturbations on the extreme field line for different values of the parameters $ \mu $ and $ \gamma $.
%%%  }
%%%  \label{tab:beta}
%%%\end{table}

\begin{table}[!t]\medskip
  \centering
  \newcolumntype{Y}{>{\centering\arraybackslash}X}
  \begin{tabularx}{\linewidth}{|c|Y|Y|Y|Y|Y|Y|}
    \hline
    $\gamma \setminus \mu$ & 1 & 2 & 3 & 4 & 5 & 6 \\ \hline
 0.5&\num{0.188684}&\num{0.332226}&\num{0.383031}&\num{0.408644}&\num{0.424036}&\num{0.434299}
 \\ \hline
   2&\bnum{0.359013}&\num{0.48633}&\num{0.52569}&\num{0.544845}&\num{0.556172}&\num{0.563656}
 \\ \hline
 6&\num{0.595414}&\num{0.681808}&\num{0.706156}&\num{0.717706}&\num{0.724454}&\num{0.72888}
 \\ \hline
  \end{tabularx}
  \caption{
    critical beta for local perturbations on the extreme field line $\psi=1$ for $f(\psi)=1-\psi^{2}$, mirror ratio $K=100$, and different values of the parameters $\mu$ and $\gamma$.
  }
  \label{tab:beta2}
\medskip
  \centering
  \newcolumntype{Y}{>{\centering\arraybackslash}X}
  \begin{tabularx}{\linewidth}{|c|Y|Y|Y|Y|Y|Y|}
    \hline
    $\gamma \setminus \mu$ & 1 & 2 & 3 & 4 & 5 & 6 \\ \hline
0.5 &\num{0.364621}&\num{0.617771}&\num{0.700717}&\num{0.740877}&\num{0.764413}&\num{0.779837}
 \\ \hline
2 &\bnum{0.646199}&\num{0.832018}&\num{0.88248}&\num{0.905453}&\num{0.918474}&\num{0.926826}
 \\ \hline
6 &\num{0.910691}&\num{0.982403}&\num{0.995413}&\num{0.999327}&\num{1.}&\num{1.}
 \\ \hline
  \end{tabularx}
  \caption{
    Same as in Table \ref{tab:beta2} for more smooth radial plasma profile, $f(\psi)=1-\psi$.
  }
  \label{tab:beta1}
\medskip
  \centering
  \newcolumntype{Y}{>{\centering\arraybackslash}X}
  \begin{tabularx}{\linewidth}{|c|Y|Y|Y|Y|Y|Y|}
    \hline
    $\gamma \setminus \mu$ & 1 & 2 & 3 & 4 & 5 & 6 \\ \hline
0.5&\num{0.0971278}&\num{0.175053}&\num{0.20366}&\num{0.218319}&\num{0.227211}&\num{0.233176}
 \\ \hline
2&\bnum{0.193448}&\num{0.26922}&\num{0.29359}&\num{0.305635}&\num{0.312819}&\num{0.317591}
 \\ \hline
6&\num{0.356438}&\num{0.418486}&\num{0.436558}&\num{0.445233}&\num{0.450334}&\num{0.453694}
 \\ \hline
  \end{tabularx}
  \caption{
    Same as in Table \ref{tab:beta2} for more sharp radial plasma profile, $f(\psi)=1-\psi^{4}$.
  }
%  \color{red} 0.193448 \num{0.193448} $0.193448$ $\num{0.193448}$\\
%  \bfseries 0.193448 \num{0.193448} $0.193448$ $\num{0.193448}$\\
%  \bfseries 0.193448 \num{0.193448} $0.193448$ $\bnum{0.193448}$\\
%
  \label{tab:beta4}
\end{table}

%По словам авторов, проделанные расчёты показывают, что профиль \eqref{4:17}, оптимальный с точки зрения желобковых мод ($\mu =1$, $\gamma =2$), становится неустойчивым относительно баллонных колебаний при $\beta_{0} >0.3$.\footnote{
%    Следует ожидать, что для линейного по $\psi$ профиля давления предельное бета будет в 2 или почти в 2 раза больше, поскольку производная $\tder{f}{\psi}$ будет в 2 раза меньше.
%} Для повышения предельных значений $\beta_{0}$  следует переходить к более крутым ступенчатым профилям ($\mu, \gamma \approx 5\div6$). Таким способом можно повысить $\beta_{\text{crit}}$ до значений $0.7\div0.8$.
%
The performed calculations show that the profile \eqref{4:17}, which is optimal for stabilizing flute modes at $ \mu = 1 $, $ \gamma = 2 $, becomes unstable with respect to ballooning vibrations at $ \beta_{0}> \num{0.359013} $. To increase the critical values of $ \beta_{0} $, Bushkova and Mirnov proposed to go to steeper axial profiles of the magnetic field described by Eq.~\eqref{4:17} with $ \mu, \gamma \approx 5 \div6 $. In this way, one can raise $ \beta_{\text{crit}} $ to the values above $0.7$.

%Чтобы изучить влияние радиального профиля давления плазмы, мы вычислили критические значения бета для более гладкого профиля давления, который описывается функцией $f(\psi)=1-\psi$, \ref{tab:beta4}), и более резкого профиля давления, который описывается функцией $f(\psi)=1-\psi^{4}$, при том же наборе значений параметров $\mu$ и $\gamma$, что и Таблице \ref{tab:beta2}. Результаты расчётов представлены соответственно в Таблицах \ref{tab:beta1} и \ref{tab:beta4}. Как и следовало ожидать по результатам обсуждения в предыдущем разделе \ref{s3}, гладкий профиль давления плазмы более устойчив относительно мелкомасштабных баллонных возмущений. Расчёт для параболического радиального профиля давления (Таблица \ref{tab:beta1}) дал очень высокие значения предельного бета, приближающиеся к единице. Большое бета получается для больших значений параметров $\mu$ и $\gamma$. Им соответствует аксиальный профиль магнитного поля, очень круто нарастающий вблизи магнитно пробки, который, вероятно, нереально сформировать в реальной ловушке. Мы сделали расчёт для таких значений $\mu$ и $\gamma$ только потому, чтобы сравнить наши результаты с результатами работы \cite{BushkovaMirnov1985BINP_103e, BushkovaMirnov1986VANT_2_19e}.

To study the effect of the radial plasma pressure profile, we calculated the critical beta values for a smoother pressure profile, described by the function $ f (\psi) = 1- \psi $, and a steeper pressure profile, described by $ f (\psi) = 1- \psi ^ {4} $, with the same set of values for the parameters $ \mu $ and $ \gamma $ as in Table \ref{tab:beta2}. The calculation results are presented respectively in Tables \ref{tab:beta1} and \ref{tab:beta4}. As would be expected from the discussion in Section~\ref{s3}, the smooth plasma pressure profile is more robust against small-scale ballooning disturbances. The calculation for a parabolic radial pressure profile (Table \ref{tab:beta1}) gave very high values of beta approaching unity. These values are obtained for large parameters $ \mu $ and $ \gamma $. They correspond to the axial profile of the magnetic field that grows very steeply near the magnetic mirror, which can hardly be formed in a real mirror trap. We made the calculation for such values of $ \mu $ and $ \gamma $ just to compare our results with the results of \cite{BushkovaMirnov1985BINP_103e, BushkovaMirnov1986VANT_2_19e}.

\section{Gas-Dynamic Trap}
\label{s5}

%Расчёты, выполненные в разделе \ref{s4} для модельного поля \eqref{4:17}, можно повторить для более реалистичных профилей вакуумного магнитного поля вдоль оси ловушки. В качестве последнего примера вычислим предельное бета для газодинамической ловушки \cite{IvanovPrikhodko2013PPCF_55_063001, IvanovPrikhodko2017PhysUsp_60_509}.

The calculations performed in Section~\ref{s4} for the model field \eqref{4:17} can be repeated for more realistic profiles of the vacuum magnetic field along the trap axis. As the last example, let us calculate the critical beta for the Gas-Dynamic Trap \cite{IvanovPrikhodko2013PPCF_55_063001, IvanovPrikhodko2017PhysUsp_60_509}.

\begin{figure}
  \centering
  \includegraphics[width=\linewidth]{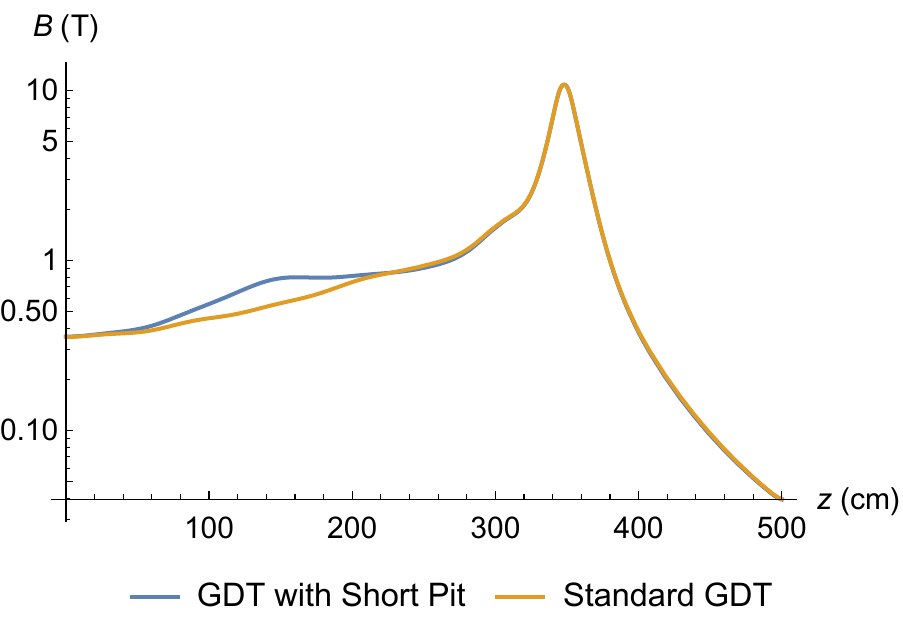}
  \caption{
    Axial profile of vacuum magnetic field in gas-dynamic trap.
  }\label{fig:B(z)-GDT}
\end{figure}
%Вакуумное магнитное поле в ГДЛ симметрично относительно экваториальной плоскости $z=0$. На рис.~\ref{fig:B(z)-GDT} показан аксиальный профиль магнитного поля в одной половине ГДЛ для двух вариантов коммутации магнитных катушек: так называемый вариант стандартной ГДЛ и вариант ГДЛ с коротким пробкотроном. Во втором варианте вблизи точки остановки быстрых ионов ($z\approx 140\,\text{см}$)  формируется неглубокий локальный минимум магнитного поля. Популяция быстрых ионов возникает при наклонной инжекции пучков нейтральных атомов под углом $45^{\circ}$ к оси ловушки.
%
The vacuum magnetic field in the GDT is symmetric about the equatorial plane $ z = 0 $. Fig.~\ref{fig:B(z)-GDT} shows the axial profile of the magnetic field in one half of the GDT for two options for switching magnetic coils: the so-called standard version of the GDT and the GDT with a short pit (short mirror cell). In the second variant, a shallow local minimum of the magnetic field is formed near the stopping point of fast ions ($ z \approx 140\,\text{cm} $). A population of fast ions arises upon oblique injection of beams of neutral atoms at the angle of $ 45^{\circ} $ to the axis of the trap.

%Из-за наличия быстрых ионов давление плазмы в ГДЛ анизотропно, т.е.\ $p_{\perp}\neq p_{\|}$, а функция $p_{\perp}$ зависит не только от магнитного потока $\psi$, но и от величины магнитного поля $B$. В центральной секции ГДЛ (т.е.\ между магнитными пробками) функция $p_{\perp}(\psi,B)$ имеет максимум вблизи точки остановки быстрых ионов, а в расширителе (т.е.\ вне центральной секции) убывает приблизительно пропорционально $B$. Вследствие всего этого оказывается, что уравнения \eqref{4:02} не удаётся решить в аналитическом виде относительно $B$ и $r$, как в случае изотропной плазмы. Таким образом, вычисление критического бета в анизотропной плазме требует существенно более сложных расчётов, чем позволяет метод, описанный в разделе \ref{s4}. Чтобы упростить себе задачу, мы игнорируем факт анизотропии и всё-таки используем метод из раздела \ref{s4}, учитывая, что давление плазмы лишь незначительно меняется вдоль силовой в области $|z|\leq 400\,\text{см}$, где магнитное поле не меньше, чем посередине центральной секции при $z=0\,\text{см}$.

Due to the presence of fast ions, the plasma pressure in the GDT is anisotropic, i.e.\  $ p_ {\perp} \neq p_{\|} $, and the function $ p_{\perp} $ depends not only on the magnetic flux $ \psi $, but also on the magnitude of the magnetic field $ B $. In the central section of the GDT (i.e., between the magnetic mirrors), function $ p_{\perp} (\psi, B) $ has a maximum near the stopping point of fast ions, and in the expander (i.e., outside the central section) it decreases roughly proportional to $ B $. As a result of all this, it turns out that Eq.~\eqref{4:02} cannot be solved analytically for $ B $ and $ r $, as in the case of an isotropic plasma. Thus, calculating the critical beta in anisotropic plasma requires much more complex calculations than the method described in Section~\ref{s4} allows. To simplify our task, we ignore the fact of anisotropy and still use the method from Section~\ref{s4}, taking into account that the plasma pressure varies only slightly along the magnetic field lines in the region $|z| \leq 400\,\text{cm}$, where the magnetic field is not less than in the middle of the central section at $ z = 0\,\text{cm} $.

%Поскольку аксиальный профиль магнитного поля вдоль оси ГДЛ жёстко задан конфигурацией магнитной системы, единственным свободным параметром, который может влиять на величину предельного бета, является координата $z_{\text{end}}$ воображаемого или реального проводящего приёмника плазмы, на котором обращается в ноль смещение $X$. Иными словами, мы вычисляли предельное значение бета, решая уравнение \eqref{4:14} с граничными условиями
%
Since the axial profile of the magnetic field along the GDT axis is rigidly set by the configuration of the magnetic system, the only free parameter that can affect the critical beta is the coordinate $ z_{\text{end}} $ of an imaginary or real conducting plasma diaphragm, on which the perturbation $X$ vanishes. In other words, we calculated the limit for beta by solving Eq.~\eqref{4:14} with the boundary conditions
    \begin{gather}
    \label{5:18}
    X(0)=1, \qquad X(z_{\text{end}})=0
    .
    \end{gather}

\begin{table}
  \centering
  \newcolumntype{Y}{>{\centering\arraybackslash}X}
  Standard GDT\\[0.5ex]
  \begin{tabularx}{\linewidth}{|c|Y|Y|Y|}
    \hline
    $f(\psi) \setminus z_{\text{end}}$ & $329.5\,\text{cm}$ & $350\,\text{cm}$ & $400\,\text{cm}$
    \\ \hline
    $1-\psi^{2}$  & \num{0.410841}  & \bnum{0.389467}  & \num{0.342571}
    \\ \hline
    $1-\psi\phantom{^{2}}$  & \num{0.718989}  & \bnum{0.689944}  & \num{0.621432}
    \\ \hline
    $1-\psi^{4}$  & \num{0.228552}  & \bnum{0.214844}  & \num{0.185939}
    \\ \hline
    \end{tabularx}
  \par
  \medskip
  GDT with Short Pit\\[0.5ex]
  \begin{tabularx}{\linewidth}{|c|Y|Y|Y|}
    \hline
    $f(\psi) \setminus z_{\text{end}}$ & $329.5\,\text{cm}$ & $350\,\text{cm}$ & $400\,\text{cm}$
    \\ \hline
    $1-\psi^{2}$  & \num{0.31577}  & \num{0.298802}  & \num{0.262309}
    \\ \hline
    $1-\psi\phantom{^{2}}$  & \num{0.579055}  & \num{0.551802}  & \num{0.490892}
    \\ \hline
    $1-\psi^{4}$  & \num{0.170379}  & \num{0.160401}  & \num{0.139505}
    \\ \hline
    \end{tabularx}
  \caption{
    Critical beta for local perturbations on the extreme field line $\psi=1$ for three radial pressure profiles $f(\psi)$ and three positions $z_{\text{end}} $ of plasma limiter or receiver.
  }
  \label{tab:GDT}
\end{table}

%Результаты расчётов представлены в Таблице \ref{tab:GDT} для двух конфигураций магнитного поля и трёх значений $z_{\text{end}}$. Координата $z_{\text{end}}=400\,\text{см}$ соответствует размещению приёмника плазмы в расширителе, а $z_{\text{end}}=350\,\text{см}$  --- в магнитной пробке. Наконец, $z_{\text{end}}=329.5\,\text{см}$ отмечает координату реально существующей проводящей диафрагмы, которая на самом деле способна играть роль стабилизатора мелкомасштабных желобковых возмущений на боковой поверхности плазменного шнура. Анализ данных в таблице показывает, что имеется весьма сильная зависимость критического значения бета от крутизны радиального профиля давления плазмы, но зависимость от координаты $z_{\text{end}}$ не кажется очень существенной.

The results of calculations are presented in Table \ref{tab:GDT} for two magnetic field configurations and three values of $ z_{\text{end}} $. The coordinate $ z_{\text{end}} = 400\,\text{cm} $ corresponds to the placement of the plasma receiver in the expander, and $ z_{\text{end}} = 350\,\text{cm} $ in the magnetic plug. Finally, $ z_{\text{end}} = 329.5\,\text{cm} $ marks the coordinate of an actually existing conducting diaphragm, which is actually capable of playing the role of a stabilizer of small-scale flute disturbances on the lateral surface of the plasma column. Analysis of the data in the table shows that there is a very strong dependence of the critical value of beta on the steepness of the radial plasma pressure profile, but the dependence on the coordinate $ z_{\text{end}} $ does not seem to be very significant. The second point to pay attention to is a noticeable decrease in critical beta in a GDT configuration with a short pit compared to the standard version of GDT. This is the expected result.

\begin{figure}
  \centering
  \includegraphics[width=\linewidth]{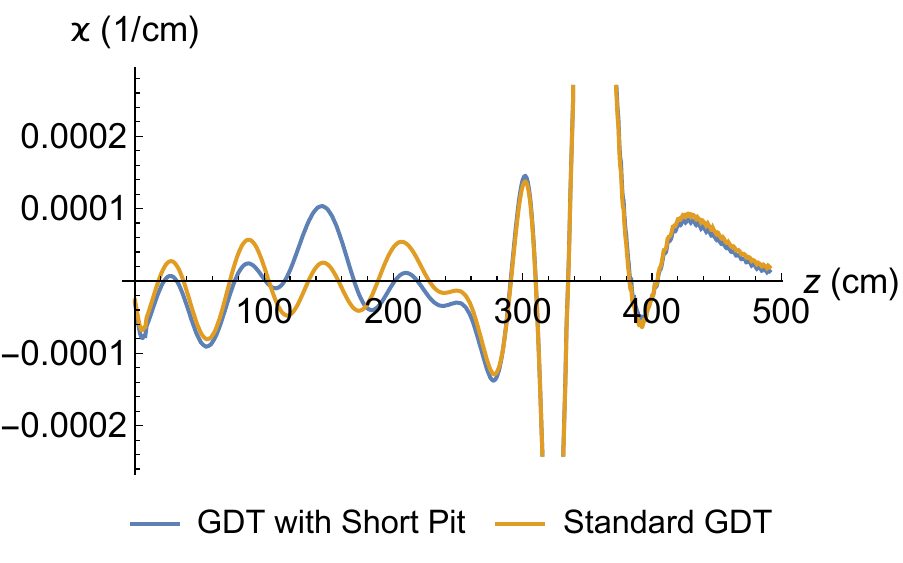}
  \caption{
    Axial profile of the vacuum magnetic field curvature in gas-dynamic trap along a field line that has unit radius at $z=0$.
  }\label{fig:kappa(z)-GDT}
\end{figure}
%К неожиданному выводу приводит сравнение результатов расчётов в таблице \ref{tab:GDT} для $z_{\text{end}}=350\,\text{см}$ (они выделены жирным шрифтом) с результатами расчёта для модельного поля \eqref{4:17}. При проектировании ГДЛ магнитная система рассчитывалась так, чтобы минимизировать дестабилизирующий вклад центральной секции в критерий желобковой устойчивости Розенблюта-Лонгмайра \cite{RosenbluthLongmire1957AnnPhys_1_120}. Такой минимум обеспечивает магнитное поле с профилем, который описывает формула \eqref{4:17} при $\mu =1$ и $\gamma =2$. Соответствующие критические значения бета в таблицах \ref{tab:beta2}, \ref{tab:beta1}, \ref{tab:beta4} также набраны жирным шрифтом. Из-за дискретной структуры магнитной системы реальное магнитное поле в ГДЛ лишь приблизительно следует формуле \eqref{4:17}. Это хорошо видно на рис.~\ref{fig:kappa(z)-GDT}, где показано, что кривизна силовой линии осциллирует, тогда как согласно формуле \eqref{4:17} она должна быть гладкой функцией. Мы ожидали, что следствием осцилляций кривизны будет снижение критического бета, но по данным таблицы \ref{tab:GDT} критическое бета оказалось несколько больше, чем в таблицах \ref{tab:beta2}, \ref{tab:beta1} и \ref{tab:beta4}. В настоящее время у нас нет объяснения этому факту. С другой стороны, заметное уменьшение критического бета в конфигурации ГДЛ с коротким пробкотроном по сравнению со стандартным вариантом ГДЛ, как это следует из анализа данных в таблице \ref{tab:GDT}, вполне соответствует нашим ожиданиям.

On the other hand, comparison of the calculation results in table \ref{tab:GDT} for $ z_{\text{end}} = 350\,\text{cm} $ (they are in bold) with the calculation results for the model field \eqref{4:17} leads to an unexpected conclusion. When designing the GDT, the magnetic field was tuned so as to minimize the destabilizing contribution of the central section to the Rosenbluth-Longmire criterion of flute stability \cite{RosenbluthLongmire1957AnnPhys_1_120}. Such a minimum is provided by a magnetic field with the profile that is described by Eq.~\eqref{4:17} for $ \mu = 1 $ and $ \gamma = 2 $. The corresponding critical beta values in tables \ref{tab:beta2}, \ref{tab:beta1}, \ref{tab:beta4} are also shown in bold. Due to the discrete structure of the magnetic system, the real magnetic field in the GDT is slightly rippled and only approximately follows the formula \eqref{4:17}. This is clearly seen in Fig.~\ref{fig:kappa(z)-GDT}, where it is shown that the curvature of the field line oscillates, whereas according to the formula \eqref{4:17} it should be a smooth function. We expected that curvature ripples would result in a decrease in critical beta, but according to table \ref{tab:GDT}, the critical beta turned out to be slightly higher than in the tables \ref{tab:beta2}, \ref{tab:beta1} and \ref{tab:beta4}. We currently have no explanation for this fact, although it worth noting that the magnetic field ripples are known to improve stability of the $m=1$ rigid ballooning mode \cite{LiKesnerLoDestro1987NF_27_1259}.

%\section{Баллонная неустойчивость в диамагнитной ловушке}

\section{Discussion}\label{s9}

We have shown that steepening of the radial plasma pressure profile leads to a decrease in the critical value of beta, above which small-scale balloon-type disturbances in a mirror trap become unstable. It means that small-scale ballooning instability leads to a smoothing of the radial plasma profile. This fact does not seem to have received due attention in the available publications. It is known that the study of the stability of the global ballooning mode leads to the opposite conclusion. For example, Ref.~\cite{LiKesnerLane1985NF_25_907} states that wall stabilization of the global ballooning mode is more effective for plasma with a hollow pressure profile.

%Мы также вычислили критические значения бета для реального магнитного поля ГДЛ в двух конфигурациях: стандартной ГДЛ и ГДЛ с короткой ямой. Для варианта стандартной ГДЛ результат вычислений оказался близок к результатам вычислений в модели, использованной ранее в статье Бушковой и Мирнова \cite{BushkovaMirnov1985BINP_103e, BushkovaMirnov1986VANT_2_19e}. В лучшем варианте критическое бета равно \num{0.718989}. Оно получается для плазмы с параболическим профилем давления по радиусу, который не был подвергнут анализу в работах \cite{BushkovaMirnov1985BINP_103e, BushkovaMirnov1986VANT_2_19e}. Максимальное бета, которое достигнуто в экспериментах на ГДЛ к настоящему времени, равно $0.6$ \cite{Simonen+2010JFE_29_558}.

We also calculated the critical beta values for the real magnetic field of the GDT in two configurations: a standard GDT and the GDT with a short well. For the version of the standard GDT, the calculation result turned out to be close to the results of calculations in the model used earlier by Bushkova and Mirnov \cite{BushkovaMirnov1985BINP_103e, BushkovaMirnov1986VANT_2_19e}. In the best case, the critical beta is \num{0.718989}. It is obtained for a plasma with a parabolic radial pressure profile, which was not analyzed in Refs.~\cite{BushkovaMirnov1985BINP_103e, BushkovaMirnov1986VANT_2_19e}. The maximum beta, which has been achieved in experiments at the GDT to date, is $ 0.6 $ \cite{Simonen+2010JFE_29_558}.

%\bigskip
\begin{acknowledgments}

    The work was financially supported by the Ministry of Education and Science of the Russian Federation.
    % (project RFMEFI61914X0003).
    %
    This study was also supported by Chinese Academy of Sciences President’s International Fellowship Initiative (PIFI) under the Grant No.~2019VMA0024 and Chinese Academy of Sciences International Partnership Program under the Grant No.~116134KYSB20200001.
    The authors are grateful to Vladimir Mirnov for valuable explanations of the method he used to solve Eq.~\eqref{4:14}, and to Dmitry Yakovlev for providing the results of calculating the magnetic field in the GDT.

\end{acknowledgments}

%\clearpage

%\providecommand \natexlab [1]{#1}%
%\providecommand \enquote  [1]{``#1''}%
%\providecommand \bibnamefont  [1]{#1}%
%\providecommand \bibfnamefont [1]{#1}%
%\providecommand \citenamefont [1]{#1}%

%\bibliographystyle{apsrev4-1}% дает ошибку
%\bibliographystyle{apsrev4-2}% дает ошибку
%\bibliographystyle{plain}
%\bibliographystyle{unsrt}
%%
%% unsrtnat дает ошибку в команде \doi при компиляции в xelatex.
%% Эту ошибку в компиляторе xelatex обещали исправить в следующем обновлении MikTeX.
%%
%%\meaning\doi
%\renewcommand{\doi}[1]{\url{https://dx.doi.org/#1}}
%\bibliographystyle{unsrtnat}
%%\bibliographystyle{ugost2008}
%\bibliography{GDMT}

\begin{thebibliography}{56}
\providecommand{\natexlab}[1]{#1}
\providecommand{\url}[1]{\texttt{#1}}
\expandafter\ifx\csname urlstyle\endcsname\relax
  \providecommand{\doi}[1]{doi: #1}\else
  \providecommand{\doi}{doi: \begingroup \urlstyle{rm}\Url}\fi

\bibitem[Ryutov and Stupakov(1980{\natexlab{a}})]{RyutovStupakov1980BINP_130e}
D.~D. Ryutov and G.~V. Stupakov.
\newblock On mhd stability of plasma in open traps with a large mirror ratio.
\newblock Technical Report 80-130, Institute of Nuclear Physics SB RAS USSR,
  Novosibirsk, 1980{\natexlab{a}}.
\newblock URL
  \url{http://www.inp.nsk.su/activity/preprints/files/1980_130.pdf}.
\newblock (in Russian).

\bibitem[Ryutov and Stupakov(1980{\natexlab{b}})]{RyutovStupakov1981IAEA_1_119}
D.D. Ryutov and G.V. Stupakov.
\newblock New results in the theory of mhd-stability and transport processes in
  ambipolar traps.
\newblock In \emph{Plasma Physics and Controlled Nuclear Fusion Research, Eight
  Conference Proceedings (Brussel, 1-10 July, 1980)}, volume~1, pages 119--132,
  Vienna, 1980{\natexlab{b}}. International Atomic Energy Agency, IAEA.

\bibitem[Bushkova and Mirnov(1985)]{BushkovaMirnov1985BINP_103e}
O.~A. Bushkova and V.~V. Mirnov.
\newblock Influence of the configuration of the magnetic field on the mhd
  stability of the gas-dynamic trap.
\newblock Technical Report 85-103, Institute of Nuclear Physics SB RAS USSR,
  Novosibirsk, 1985.
\newblock URL
  \url{http://www.inp.nsk.su/activity/preprints/files/1985_103.pdf}.
\newblock (in Russian).

\bibitem[Bushkova and Mirnov(1986)]{BushkovaMirnov1986VANT_2_19e}
O.~A. Bushkova and V.~V. Mirnov.
\newblock Influence of the configuration of the magnetic field on the mhd
  stability of the gas-dynamic trap.
\newblock \emph{Questions of atomic science and technology, ser. Thermonuclear
  fusion}, \penalty0 (2):\penalty0 19--24, 1986.
\newblock (in Russian).

\bibitem[Newcomb(1981)]{Newcomb1981JPP_26_529}
W.~A. Newcomb.
\newblock Equilibrium and stability of collisionless systems in the paraxial
  limit.
\newblock \emph{J. Plasma Physics}, 26\penalty0 (3):\penalty0 529--584, 1981.

\bibitem[D’Ippolito and Myra(1981)]{DIppolitoMyra1981PF_24_2265}
D.~A. D’Ippolito and J.~R. Myra.
\newblock Strongly‐localized ballooning modes in an axisymmetric tandem
  mirror.
\newblock \emph{The Physics of Fluids}, 24\penalty0 (12):\penalty0 2265--2269,
  1981.
\newblock \doi{10.1063/1.863345}.
\newblock URL \url{https://aip.scitation.org/doi/abs/10.1063/1.863345}.

\bibitem[D'Ippolito et~al.(1982)D'Ippolito, Myra, and
  Ogden]{DIppolitoMyraOgden1982NF_24_707}
D.~A. D'Ippolito, J.~R. Myra, and J.~M. Ogden.
\newblock High-$m$ ballooning stability of an axisymmetric e-ring-stabilized
  tandem mirror.
\newblock \emph{Plasma Physics}, 24\penalty0 (7):\penalty0 707--730, jul 1982.
\newblock \doi{10.1088/0032-1028/24/7/001}.
\newblock URL \url{https://doi.org/10.1088/0032-1028/24/7/001}.

\bibitem[D’Ippolito et~al.(1982)D’Ippolito, Hafizi, and
  Myra]{DIppolitoHafiziMyra1982PF_25_2223}
D.~A. D’Ippolito, B.~Hafizi, and J.~R. Myra.
\newblock Ideal magnetohydrodynamic stability of axisymmetric mirrors.
\newblock \emph{The Physics of Fluids}, 25\penalty0 (12):\penalty0 2223--2230,
  1982.
\newblock \doi{10.1063/1.863962}.
\newblock URL \url{https://aip.scitation.org/doi/abs/10.1063/1.863962}.

\bibitem[Bernstein et~al.(1958)Bernstein, Frieman, Kruskal, and
  Kulsrud]{Bernstein+1958ProcRSos_17_244}
I.~B. Bernstein, E.~A. Frieman, M.~D. Kruskal, and R.~M. Kulsrud.
\newblock An energy principle for hydromagnetic stability problems.
\newblock \emph{Proc.\ R.\ Soc.\ London, Ser.\ A}, 244:\penalty0 17, 1958.
\newblock \doi{10.1098/rspa.1958.0023}.
\newblock URL
  \url{http://rspa.royalsocietypublishing.org/content/244/1236/17.short}.

\bibitem[Kaiser and Pearlstein(1983)]{KaiserPearlstein1983PhysFluids_26_3053}
Thomas~B. Kaiser and L.~Donald Pearlstein.
\newblock Ballooning modes in quadrupole tandem mirrors.
\newblock \emph{The Physics of Fluids}, 26\penalty0 (10):\penalty0 3053--3065,
  1983.
\newblock \doi{10.1063/1.864029}.
\newblock URL \url{https://aip.scitation.org/doi/abs/10.1063/1.864029}.

\bibitem[Close and Lichtenberg(1989)]{CloseLichtenberg1989PFB_1_629}
R.~A. Close and A.~J. Lichtenberg.
\newblock Ballooning modes in an axisymmetric mirror machine.
\newblock \emph{Physics of Fluids B: Plasma Physics}, 1\penalty0 (3):\penalty0
  629--634, 1989.
\newblock \doi{10.1063/1.859122}.
\newblock URL \url{https://doi.org/10.1063/1.859122}.

\bibitem[Tsidulko(1992)]{Tsidulko1992BINP_92-10e}
Y.~A Tsidulko.
\newblock Resistive balloon mode in a gas-dynamic trap.
\newblock Preprint INP SB RAS 92-10, INP SB AS USSR, Novosibirsk, 1992.
\newblock URL
  \url{http://www.inp.nsk.su/activity/preprints/files/1992_010.pdf}.
\newblock (in Russian).

\bibitem[Bilikmen et~al.(1997)Bilikmen, Mirnov, and
  Oke]{BilikmenMirnovOke1997NF_37_973}
S.~Bilikmen, V.~V. Mirnov, and G.~Oke.
\newblock Localized ballooning modes in two component gas dynamic trap.
\newblock \emph{Nuclear Fusion}, 37\penalty0 (7):\penalty0 973--983, jul 1997.
\newblock \doi{10.1088/0029-5515/37/7/i06}.
\newblock URL \url{https://doi.org/10.1088/0029-5515/37/7/i06}.

\bibitem[Arsenin(1983)]{Arsenin1983JETPhLett_37_637}
V.~V. Arsenin.
\newblock Mhd stability of a low-pressure plasma in an axisymmetric open system
  with an alternating-sign curvature.
\newblock \emph{JETP Lett.}, 37\penalty0 (11):\penalty0 637--640, 1983.

\bibitem[Arsenin and Kuyanov(2001)]{ArseninKuyanov2001FT_39_175}
Vladimir~V. Arsenin and Alexey~Yu. Kuyanov.
\newblock Non-paraxial plasma equilibria in axisymmetric mirrors.
\newblock \emph{Fusion Technology}, 39\penalty0 (1T):\penalty0 175--178, 2001.
\newblock \doi{10.13182/FST01-A11963435}.
\newblock URL \url{https://doi.org/10.13182/FST01-A11963435}.

\bibitem[Arsenin et~al.(2005)Arsenin, Zvonkov, and
  Skovoroda]{ArseninZvonkovSkovoroda2005PPR_31_3}
V.~V. Arsenin, A.~V. Zvonkov, and A.~A. Skovoroda.
\newblock Stabilization of ballooning modes by nonparaxial cells.
\newblock \emph{Plasma Phys. Rep.}, 31:\penalty0 3--13, 2005.
\newblock \doi{10.1134/1.1856704}.
\newblock URL \url{https://doi.org/10.1134/1.1856704}.

\bibitem[Arsenin(2008)]{Arsenin2008PPR_34_349}
V.~V. Arsenin.
\newblock Mhd stability of a finite-pressure plasma in axisymmetric
  configurations of the poloidal magnetic field.
\newblock \emph{Plasma Phys. Rep.}, 34:\penalty0 349, 2008.
\newblock \doi{10.1134/S1063780X08050012}.
\newblock URL \url{https://doi.org/10.1134/S1063780X08050012}.

\bibitem[Arsenin and Terekhin(2008)]{ArseninTerekhin2008PPR_34_895}
V.~V. Arsenin and P.~N. Terekhin.
\newblock Mhd stability condition of an anisotropic-pressure plasma in axially
  symmetric confinement systems formed by a poloidal field.
\newblock \emph{Plasma Phys. Rep.}, 34:\penalty0 895, 2008.
\newblock \doi{10.1134/S1063780X08110020}.
\newblock URL \url{https://doi.org/10.1134/S1063780X08110020}.

\bibitem[Arsenin and Terekhin(2011)]{ArseninTerekhin2011PPR_37_723}
V.~V. Arsenin and P.~N. Terekhin.
\newblock Plasma equilibrium in axisymmetric poloidal magnetic field
  configurations in flux coordinates.
\newblock \emph{Plasma Phys. Rep.}, 37:\penalty0 723, 2011.
\newblock \doi{10.1134/S1063780X11070038}.
\newblock URL \url{https://doi.org/10.1134/S1063780X11070038}.

\bibitem[Rosenbluth et~al.(1962)Rosenbluth, Krall, and
  Rostoker]{Rosenbluth+1962NFSuppl_1_143}
M.N. Rosenbluth, N.A. Krall, and N.~Rostoker.
\newblock Finite larmor radius stabilization of ``weakly'' unstable confined
  plasmas.
\newblock \emph{Nuclear Fusion}, Suppl., Part 1:\penalty0 143--150, 1962.
\newblock URL \url{http://www-naweb.iaea.org/napc/physics/FEC/1961.pdf}.

\bibitem[Roberts and Taylor(1962)]{RobertsTaylor1962PhysRevLett_8_197}
K.~V. Roberts and J.~B. Taylor.
\newblock Magnetohydrodynamic equations for finite larmor radius.
\newblock \emph{Phys. Rev. Lett.}, 8:\penalty0 197--198, Mar 1962.
\newblock \doi{10.1103/PhysRevLett.8.197}.
\newblock URL \url{https://link.aps.org/doi/10.1103/PhysRevLett.8.197}.

\bibitem[Rudakov(1962)]{Rudakov1962NF_2_107}
L.~I. Rudakov.
\newblock Influence of the viscosity of plasma in a magnetic field on plasma
  stability.
\newblock \emph{Nuclear Fusion}, 2\penalty0 (1-2):\penalty0 107--108, jan 1962.
\newblock \doi{10.1088/0029-5515/2/1-2/016}.
\newblock URL \url{https://doi.org/10.1088/0029-5515/2/1-2/016}.

\bibitem[D’Ippolito et~al.(1981)D’Ippolito, Francis, Myra, and
  Tang]{DIppolitoFrancisMyraTang1981PF_24_2270}
D.~A. D’Ippolito, G.~L. Francis, J.~R. Myra, and W.~M. Tang.
\newblock Finite larmor radius stabilization of ballooning modes in an
  axisymmetric tandem mirror.
\newblock \emph{The Physics of Fluids}, 24\penalty0 (12):\penalty0 2270--2273,
  1981.
\newblock \doi{10.1063/1.863346}.
\newblock URL \url{https://aip.scitation.org/doi/abs/10.1063/1.863346}.

\bibitem[Kaiser et~al.(1983)Kaiser, Nevins, and
  Pearlstein]{KaiserNevinsPearlstein1983PF_26_351}
Thomas~B. Kaiser, William~McCay Nevins, and L.~Donald Pearlstein.
\newblock Rigid ballooning modes in tandem mirrors.
\newblock \emph{The Physics of Fluids}, 26\penalty0 (2):\penalty0 351--353,
  1983.
\newblock \doi{10.1063/1.864170}.
\newblock URL \url{https://aip.scitation.org/doi/abs/10.1063/1.864170}.

\bibitem[Berk et~al.(1985)Berk, Horton, Rosenbluth, Wong, Kesner, Lane, Jr.,
  Tsang, Lee, Hafizi, Byers, Cohen, Hammer, Nevins, Kaiser, Lodestro,
  Pearlstein, Smith, Ramachandran, and Tang]{Berk+1985PPCNFR_2_321}
H.L. Berk, C.~W.~Jr. Horton, M.N. Rosenbluth, H.~V. Wong, J.~Kesner, B.~Lane,
  T.M.~Antonsen Jr., K.T. Tsang, X.S. Lee, B.~Hafizi, J.A. Byers, R.H. Cohen,
  J.H. Hammer, W.M. Nevins, T.B. Kaiser, L.~Lodestro, L.D. Pearlstein, G.R.
  Smith, R.~Ramachandran, and W.M. Tang.
\newblock Stabilization of an axisymmetric mirror cell and trapped particle
  modes.
\newblock In \emph{Plasma physics and controlled nuclear fusion research 1984},
  volume~2, pages 321--335, Vienna (Austria), 1985. IAEA, Nuclear Fusion,
  Suppl.

\bibitem[Berk et~al.(1984)Berk, Rosenbluth, Wong, and
  Antonsen]{Berk+1984PF_27_2705}
H.~L. Berk, M.~N. Rosenbluth, H.~Vernon Wong, and Thomas~M. Antonsen.
\newblock Stabilization of an axisymmetric tandem mirror cell by a hot plasma
  component.
\newblock \emph{The Physics of Fluids}, 27\penalty0 (11):\penalty0 2705--2710,
  1984.
\newblock \doi{10.1063/1.864574}.
\newblock URL \url{https://aip.scitation.org/doi/abs/10.1063/1.864574}.

\bibitem[D’Ippolito and Hafizi(1981)]{DIppolitoHafizi1981PF_24_2274}
D.~A. D’Ippolito and B.~Hafizi.
\newblock Low‐$m$ ballooning stability of an axisymmetric sharp‐boundary
  tandem mirror.
\newblock \emph{The Physics of Fluids}, 24\penalty0 (12):\penalty0 2274--2279,
  1981.
\newblock \doi{10.1063/1.863347}.
\newblock URL \url{https://aip.scitation.org/doi/abs/10.1063/1.863347}.

\bibitem[D’Ippolito and Myra(1984)]{DIppolitoMyra1984PF_27_2256}
D.~A. D’Ippolito and J.~R. Myra.
\newblock Stability of mirrors with inverted pressure profiles.
\newblock \emph{The Physics of Fluids}, 27\penalty0 (9):\penalty0 2256--2263,
  1984.
\newblock \doi{10.1063/1.864880}.
\newblock URL \url{https://aip.scitation.org/doi/abs/10.1063/1.864880}.

\bibitem[Kaiser and Pearlstein(1985)]{KaiserPearlstein1985PhysFluids_28_1003}
T.~B. Kaiser and L.~Donald Pearlstein.
\newblock Finite larmor radius and wall effects on the $m=1$ ballooning mode at
  arbitrary beta in axisymmetric tandem mirrors.
\newblock \emph{The Physics of Fluids}, 28\penalty0 (3):\penalty0 1003--1005,
  1985.
\newblock \doi{10.1063/1.865092}.
\newblock URL \url{https://aip.scitation.org/doi/abs/10.1063/1.865092}.

\bibitem[Kesner(1985)]{Kesner1985NF_25_275}
J.~Kesner.
\newblock Axisymmetric, wall-stabilized tandem mirrors.
\newblock \emph{Nuclear Fusion}, 25\penalty0 (3):\penalty0 275--282, mar 1985.
\newblock \doi{10.1088/0029-5515/25/3/004}.
\newblock URL \url{https://doi.org/10.1088/0029-5515/25/3/004}.

\bibitem[Li et~al.(1985)Li, Kesner, and Lane]{LiKesnerLane1985NF_25_907}
Xing-Zhong Li, J.~Kesner, and B.~Lane.
\newblock Conducting-wall and pressure profile effect on {MHD} stabilization of
  axisymmetric mirror.
\newblock \emph{Nuclear Fusion}, 25\penalty0 (8):\penalty0 907--917, aug 1985.
\newblock \doi{10.1088/0029-5515/25/8/004}.
\newblock URL \url{https://doi.org/10.1088/0029-5515/25/8/004}.

\bibitem[Li et~al.(1987{\natexlab{a}})Li, Kesner, and
  Lane]{LiKesnerLane1987NF_27_101}
Xing~Zhong Li, J.~Kesner, and B.~Lane.
\newblock {MHD} stabilization of a high beta mirror plasma partially enclosed
  by a conducting wall.
\newblock \emph{Nuclear Fusion}, 27\penalty0 (1):\penalty0 101--107, jan
  1987{\natexlab{a}}.
\newblock \doi{10.1088/0029-5515/27/1/008}.
\newblock URL \url{https://doi.org/10.1088/0029-5515/27/1/008}.

\bibitem[Li et~al.(1987{\natexlab{b}})Li, Kesner, and
  LoDestro]{LiKesnerLoDestro1987NF_27_1259}
Zing-Zhong Li, J.~Kesner, and L.L. LoDestro.
\newblock Wall stabilized high beta mirror plasma in a rippled magnetic field.
\newblock \emph{Nuclear Fusion}, 27\penalty0 (8):\penalty0 1259--1266, aug
  1987{\natexlab{b}}.
\newblock \doi{10.1088/0029-5515/27/8/007}.
\newblock URL \url{https://doi.org/10.1088/0029-5515/27/8/007}.

\bibitem[LoDestro(1986)]{LoDestro1986PF_29_2329}
L.~L. LoDestro.
\newblock The rigid ballooning mode in finite‐beta axisymmetric plasmas with
  diffuse pressure profiles.
\newblock \emph{The Physics of Fluids}, 29\penalty0 (7):\penalty0 2329--2332,
  1986.
\newblock \doi{10.1063/1.865572}.
\newblock URL \url{https://aip.scitation.org/doi/abs/10.1063/1.865572}.

\bibitem[Arsenin and Kuyanov(1996)]{ArseninKuyanov1996PPR_22_638}
V.~V. Arsenin and A.~Yu. Kuyanov.
\newblock Stabilization of $m=1$ mhd modes in axisymmetric mirror devices with
  $\beta\sim1$ plasma.
\newblock \emph{Plasma Physics Reports}, 22\penalty0 (8):\penalty0 638--642,
  1996.
\newblock \doi{https://doi.org/10.1134/1.952334}.

\bibitem[Snyder et~al.(2002)Snyder, Wilson, Ferron, Lao, Leonard, Osborne,
  Turnbull, Mossessian, Murakami, and Xu]{Snyder+2002PoP_9_2037}
P.~B. Snyder, H.~R. Wilson, J.~R. Ferron, L.~L. Lao, A.~W. Leonard, T.~H.
  Osborne, A.~D. Turnbull, D.~Mossessian, M.~Murakami, and X.~Q. Xu.
\newblock Edge localized modes and the pedestal: A model based on coupled
  peeling–ballooning modes.
\newblock \emph{Physics of Plasmas}, 9\penalty0 (5):\penalty0 2037--2043, 2002.
\newblock \doi{10.1063/1.1449463}.
\newblock URL \url{https://doi.org/10.1063/1.1449463}.

\bibitem[Halpern et~al.(2013)Halpern, Jolliet, Loizu, Mosetto, and
  Ricci]{Halpern+2013PoP_20_052306}
Federico~D. Halpern, Sebastien Jolliet, Joaquim Loizu, Annamaria Mosetto, and
  Paolo Ricci.
\newblock Ideal ballooning modes in the tokamak scrape-off layer.
\newblock \emph{Physics of Plasmas}, 20\penalty0 (5):\penalty0 052306, 2013.
\newblock \doi{10.1063/1.4807333}.
\newblock URL \url{https://doi.org/10.1063/1.4807333}.

\bibitem[Eich et~al.(2018)Eich, Goldston, Kallenbach, Sieglin, Sun, and
  and]{Eich+2018NF_58_034001}
T.~Eich, R.J. Goldston, A.~Kallenbach, B.~Sieglin, H.J. Sun, and and.
\newblock Correlation of the tokamak h-mode density limit with ballooning
  stability at the separatrix.
\newblock \emph{Nuclear Fusion}, 58\penalty0 (3):\penalty0 034001, jan 2018.
\newblock \doi{10.1088/1741-4326/aaa340}.
\newblock URL \url{https://doi.org/10.1088/1741-4326/aaa340}.

\bibitem[Ongena et~al.(2016)Ongena, Koch, and
  Wolf]{Ongena+2016NaturePhysics_12_398}
J.~Ongena, R.~Koch, and H.~Wolf, R.~Zohm.
\newblock Magnetic-confinement fusion.
\newblock \emph{Nature Physics}, 12\penalty0 (5):\penalty0 398--410, 2016.
\newblock \doi{10.1038/nphys3745}.
\newblock URL \url{https://doi.org/10.1038/nphys3745}.

\bibitem[Ivanov et~al.(2003)Ivanov, Anikeev, Bagryansky, Deichuli, Korepanov,
  Lizunov, Maximov, Murakhtin, Savkin, Den~Hartog, Fiksel, and
  Noack]{Ivanov+PhysRevLett_90_105002}
A.~A. Ivanov, A.~V. Anikeev, P.~A. Bagryansky, P.~P. Deichuli, S.~A. Korepanov,
  A.~A. Lizunov, V.~V. Maximov, S.~V. Murakhtin, V.~Ya. Savkin, D.~J.
  Den~Hartog, G.~Fiksel, and K.~Noack.
\newblock Experimental evidence of high-beta plasma confinement in an axially
  symmetric gas dynamic trap.
\newblock \emph{Phys. Rev. Lett.}, 90:\penalty0 105002, Mar 2003.
\newblock \doi{10.1103/PhysRevLett.90.105002}.
\newblock URL \url{https://link.aps.org/doi/10.1103/PhysRevLett.90.105002}.

\bibitem[Simonen et~al.(2010)Simonen, Anikeev, Bagryansky, Beklemishev, Ivanov,
  Lizunov, Maximov, Prikhodko, and Tsidulko]{Simonen+2010JFE_29_558}
T.~C. Simonen, A.~Anikeev, P.~Bagryansky, A.~Beklemishev, A.~Ivanov,
  A.~Lizunov, V.~Maximov, V.~Prikhodko, and Yu. Tsidulko.
\newblock High beta experiments in the gdt axisymmetric magnetic mirror.
\newblock \emph{Journal of Fusion Energy}, 29\penalty0 (6):\penalty0 558--560,
  Dec 2010.
\newblock ISSN 1572-9591.
\newblock \doi{10.1007/s10894-010-9342-7}.
\newblock URL \url{10.1007/s10894-010-9342-7}.

\bibitem[Bagryansky et~al.(2011)Bagryansky, Anikeev, Beklemishev, Donin,
  Ivanov, Korzhavina, Kovalenko, Kruglyakov, Lizunov, Maximov, Murakhtin,
  Prikhodko, Pinzhenin, Pushkareva, Savkin, and
  Zaytsev]{Bagryansky+2011FST_59_31}
P.~A. Bagryansky, A.~V. Anikeev, A.~D. Beklemishev, A.~S. Donin, A.~A. Ivanov,
  M.~S. Korzhavina, Yu.~V. Kovalenko, E.~P. Kruglyakov, A.~A. Lizunov, V.~V.
  Maximov, S.~V. Murakhtin, V.~V. Prikhodko, E.~I. Pinzhenin, A.~N. Pushkareva,
  V.~Ya. Savkin, and K.~V. Zaytsev.
\newblock Confinement of hot ion plasma with $\beta=0.6$ in the gas dynamic
  trap.
\newblock \emph{Fusion Science and Technology}, 59\penalty0 (1T):\penalty0
  31--35, 2011.
\newblock \doi{10.13182/FST11-A11568}.
\newblock URL \url{https://doi.org/10.13182/FST11-A11568}.

\bibitem[Bagryansky et~al.(2015{\natexlab{a}})Bagryansky, Shalashov,
  Gospodchikov, Lizunov, Maximov, Prikhodko, Soldatkina, Solomakhin, and
  Yakovlev]{Bagryansky+2015PhysRevLett_114_205001}
P.~A. Bagryansky, A.~G. Shalashov, E.~D. Gospodchikov, A.~A. Lizunov, V.~V.
  Maximov, V.~V. Prikhodko, E.~I. Soldatkina, A.~L. Solomakhin, and D.~V.
  Yakovlev.
\newblock Threefold increase of the bulk electron temperature of plasma
  discharges in a magnetic mirror device.
\newblock \emph{Phys. Rev. Lett.}, 114:\penalty0 205001, May
  2015{\natexlab{a}}.
\newblock \doi{10.1103/PhysRevLett.114.205001}.
\newblock URL \url{https://link.aps.org/doi/10.1103/PhysRevLett.114.205001}.

\bibitem[Bagryansky et~al.(2015{\natexlab{b}})Bagryansky, Anikeev, Denisov,
  Gospodchikov, Ivanov, Lizunov, Kovalenko, Malygin, Maximov, Korobeinikova,
  Murakhtin, Pinzhenin, Prikhodko, Savkin, Shalashov, Smolyakova, Soldatkina,
  Solomakhin, Yakovlev, and Zaytsev]{Bagryansky+2015NF_55_053009}
P.~A. Bagryansky, A.~V. Anikeev, G.~G. Denisov, E.~D. Gospodchikov, A.~A.
  Ivanov, A.~A. Lizunov, Yu.~V. Kovalenko, V.~I. Malygin, V.~V. Maximov, O.~A.
  Korobeinikova, S.~V. Murakhtin, E.~I. Pinzhenin, V.~V. Prikhodko, V.~Ya.
  Savkin, A.~G. Shalashov, O.~B. Smolyakova, E.~I. Soldatkina, A.~L.
  Solomakhin, D.~V. Yakovlev, and K.~V. Zaytsev.
\newblock Overview of {ECR} plasma heating experiment in the {GDT} magnetic
  mirror.
\newblock \emph{Nuclear Fusion}, 55\penalty0 (5):\penalty0 053009, apr
  2015{\natexlab{b}}.
\newblock \doi{10.1088/0029-5515/55/5/053009}.
\newblock URL \url{https://doi.org/10.1088/0029-5515/55/5/053009}.

\bibitem[Bagryansky et~al.(2016{\natexlab{a}})Bagryansky, Akhmetov,
  Chernoshtanov, Deichuli, Ivanov, Lizunov, Maximov, Mishagin, Murakhtin,
  Pinzhenin, Pikhodko, Sorokin, and
  Oreshonok]{Bagryansky+2016AIPConfProc_030015}
P.~A. Bagryansky, T.~D. Akhmetov, I.~S. Chernoshtanov, P.~P. Deichuli, A.~A.
  Ivanov, A.~A. Lizunov, V.~V. Maximov, V.~V. Mishagin, S.~V. Murakhtin, E.~I.
  Pinzhenin, V.~V. Pikhodko, A.~V. Sorokin, and V.~V. Oreshonok.
\newblock Status of the experiment on magnetic field reversal at binp.
\newblock \emph{AIP Conference Proceedings}, 1771\penalty0 (1):\penalty0
  030015, 2016{\natexlab{a}}.
\newblock \doi{10.1063/1.4964171}.
\newblock URL \url{http://aip.scitation.org/doi/abs/10.1063/1.4964171}.

\bibitem[Bagryansky et~al.(2016{\natexlab{b}})Bagryansky, Anikeev, Anikeev,
  Dunaevsky, Gospodchikov, Ivanov, Lizunov, Korobeynikova, Korzhavina,
  Kovalenko, Maximov, Murakhtin, Pinzhenin, Prikhodko, Savkin, Shalashov,
  Soldatkina, Solomakhin, Yakovlev, Yushmanov, and
  Zaytsev]{Bagryansky+2016AIPCP_1771_020003}
P.~A. Bagryansky, A.~V. Anikeev, M.~A. Anikeev, A.~Dunaevsky, E.~D.
  Gospodchikov, A.~A. Ivanov, A.~A. Lizunov, O.~A. Korobeynikova, M.~S.
  Korzhavina, Yu.~V. Kovalenko, V.~V. Maximov, S.~V. Murakhtin, E.~I.
  Pinzhenin, V.~V. Prikhodko, V.~Ya. Savkin, A.~G. Shalashov, E.~I. Soldatkina,
  A.~L. Solomakhin, D.~V. Yakovlev, P.~Yushmanov, and K.~V. Zaytsev.
\newblock Recent progress of plasma confinement and heating studies in the gas
  dynamic trap.
\newblock \emph{AIP Conference Proceedings}, 1771\penalty0 (1):\penalty0
  020003, 2016{\natexlab{b}}.
\newblock \doi{10.1063/1.4964156}.
\newblock URL \url{https://aip.scitation.org/doi/abs/10.1063/1.4964156}.

\bibitem[Yakovlev et~al.(2018)Yakovlev, Shalashov, Gospodchikov, Maximov,
  Prikhodko, Savkin, Soldatkina, Solomakhin, and
  Bagryansky]{Yakovlev+2018NF_58_094001}
D.~V. Yakovlev, A.~G. Shalashov, E.~D. Gospodchikov, V.~V. Maximov, V.~V.
  Prikhodko, V.~Ya. Savkin, E.~I. Soldatkina, A.~L. Solomakhin, and P.~A.
  Bagryansky.
\newblock Stable confinement of high-electron-temperature plasmas in the {GDT}
  experiment.
\newblock \emph{Nuclear Fusion}, 58\penalty0 (9):\penalty0 094001, jul 2018.
\newblock \doi{10.1088/1741-4326/aacb88}.
\newblock URL \url{https://doi.org/10.1088/1741-4326/aacb88}.

\bibitem[Bagryansky et~al.(2019)Bagryansky, Beklemishev, and
  Postupaev]{Bagryansky+2019JFE_38_162}
P.~A. Bagryansky, A.~D. Beklemishev, and V.~V. Postupaev.
\newblock Encouraging results and new ideas for fusion in linear traps.
\newblock \emph{Journal of Fusion Energy}, 38\penalty0 (1):\penalty0 162--181,
  Feb 2019.
\newblock ISSN 1572-9591.
\newblock \doi{10.1007/s10894-018-0174-1}.
\newblock URL \url{https://doi.org/10.1007/s10894-018-0174-1}.

\bibitem[Beklemishev(2016)]{Beklemishev2016PoP_23_082506}
A.~D. Beklemishev.
\newblock Diamagnetic “bubble” equilibria in linear traps.
\newblock \emph{Physics of Plasmas}, 23\penalty0 (8):\penalty0 082506, 2016.
\newblock \doi{10.1063/1.4960129}.
\newblock URL \url{https://doi.org/10.1063/1.4960129}.

\bibitem[{Granetzny} et~al.(2018){Granetzny}, {Anderson}, {Clark}, {Green},
  {Schmitz}, and {Forest}]{Granetzny+2018APS_CP11_150}
Marcel {Granetzny}, Jay {Anderson}, Mike {Clark}, Jonathan {Green}, Oliver
  {Schmitz}, and Cary {Forest}.
\newblock {Device overview and first results from the gas dynamic trap
  prototype}.
\newblock In \emph{APS Division of Plasma Physics Meeting Abstracts}, volume
  2018 of \emph{APS Meeting Abstracts}, page CP11.150, January 2018.

\bibitem[Bagryansky et~al.(2020)Bagryansky, Chen, Kotelnikov, Yakovlev,
  Prikhodko, Zeng, Bai, Yu, Ivanov, and
  Wu]{Bagryansky+2020NuclFusion_60_036005}
P.~A. Bagryansky, Z.~Chen, I.~A. Kotelnikov, D.~V. Yakovlev, V.~V. Prikhodko,
  Q.~Zeng, Y.~Bai, J.~Yu, A.A. Ivanov, and Y.~Wu.
\newblock Development strategy for steady-state fusion volumetric neutron
  source based on the gas-dynamic trap.
\newblock \emph{Nuclear Fusion}, 60\penalty0 (3):\penalty0 036005, jan 2020.
\newblock \doi{10.1088/1741-4326/ab668d}.
\newblock URL \url{https://doi.org/10.1088/1741-4326/ab668d}.

\bibitem[Ryutov et~al.(2011)Ryutov, Berk, Cohen, Molvik, and
  Simonen]{Ryutov+2011PoP_18_092301}
D.~D. Ryutov, H.~L. Berk, B.~I. Cohen, A.~W. Molvik, and T.~C. Simonen.
\newblock Magneto-hydrodynamically stable axisymmetric mirrors.
\newblock \emph{Physics of Plasmas}, 18\penalty0 (9):\penalty0 092301, 2011.
\newblock \doi{10.1063/1.3624763}.
\newblock URL \url{https://doi.org/10.1063/1.3624763}.

\bibitem[Kotelnikov(2021)]{Kotelnikov2020V2e}
I.~A. Kotelnikov.
\newblock \emph{Magnetic hydrodynamics. Lectures and Problems}, volume~2.
\newblock Lanbook, 3 edition, 2021.
\newblock (in Russian).

\bibitem[Ivanov and Prikhodko(2013)]{IvanovPrikhodko2013PPCF_55_063001}
A.~A. Ivanov and V.~V. Prikhodko.
\newblock Gas-dynamic trap: an overview of the concept and experimental
  results.
\newblock \emph{Plasma Physics and Controlled Fusion}, 55\penalty0
  (6):\penalty0 063001, 2013.
\newblock URL \url{http://stacks.iop.org/0741-3335/55/i=6/a=063001}.

\bibitem[Ivanov and Prikhodko(2017)]{IvanovPrikhodko2017PhysUsp_60_509}
A.~A. Ivanov and V.~V. Prikhodko.
\newblock Gas dynamic trap: experimental results and future prospects.
\newblock \emph{Phys. Usp.}, 60\penalty0 (5):\penalty0 509--533, 2017.
\newblock \doi{10.3367/UFNe.2016.09.037967}.
\newblock URL \url{https://ufn.ru/en/articles/2017/5/d/}.

\bibitem[Rosenbluth and Longmire(1957)]{RosenbluthLongmire1957AnnPhys_1_120}
Marshall~N. Rosenbluth and C.~L. Longmire.
\newblock Stability of plasmas confined by magnetic fields.
\newblock \emph{Annals of Physics}, 1\penalty0 (2):\penalty0 120--140, 1957.
\newblock ISSN 0003-4916.
\newblock \doi{https://doi.org/10.1016/0003-4916(57)90055-6}.
\newblock URL
  \url{http://www.sciencedirect.com/science/article/pii/0003491657900556}.

\end{thebibliography}

\end{document}